\documentclass[useAMS,usenatbib]{mn2e}

\newcommand{\kms}{ km s$^{-1}$}
\newcommand{\ang}{\mbox{\AA} }
\usepackage{graphicx}
\usepackage{times}
\usepackage{colordvi}
\usepackage{epsfig}
\usepackage{lscape}
\usepackage{rotate}
\begin{document}

\title[Metals in  sub-Damped Lyman-$\alpha$ Absorbers at $z < 1.5$]{New abundance determinations in $z < 1.5$ QSO absorbers: seven sub-DLAs and one DLA.}
\author[J. D. Meiring, J.T. Lauroesch et al.]{Joseph D. Meiring$^{1}$, 
James T. Lauroesch$^{2,3}$, 
Varsha P. Kulkarni$^{1}$, 
Celine P\'eroux$^{4}$,
\newauthor  Pushpa Khare$^{5}$, Donald G. York$^{6,7}$, $\&$ Arlin P. S. Crotts$^{8}$\\ 
$^{1}$Department of Physics and Astronomy, University of South Carolina, Columbia, SC 29208, USA \\
$^{2}$Department of Physics and Astronomy, Northwestern University, Evanston, IL 60208, USA \\ 
$^{3}$Department of Physics and Astronomy, University of Louisville, Louisville, Ky 40292 USA\\
$^{4}$European Southern Observatory, Garching-bei-Munchen, Germany \\
$^{5}$Department of Physics, Utkal University, Bhubaneswar, 751004, India \\
$^{6}$Department of Astronomy and Astrophysics, University of Chicago, Chicago, IL 60637, USA \\ 
$^{7}$Enrico Fermi Institute, University of Chicago, Chicago, IL 60637, USA \\
$^{8}$Department of Astronomy, Columbia University, New York, NY 10027, USA\\}

\date{Accepted ... Received ...; in original form ...}

\pagerange{\pageref{firstpage}--\pageref{lastpage}} \pubyear{}

\maketitle

\label{firstpage}

\begin{abstract}
We present chemical abundance measurements from high resolution observations of 7 sub-damped Lyman-$\alpha$ absorbers and 
1 damped Lyman-$\alpha$ system at $z<1.5$. Three of these objects have high metallicity, with near or super-solar Zn abundance.
Grids of Cloudy models for each system were constructed to look for possible ionization effects in these systems. 
For the systems in which we could constrain the ionization parameter, we find that the ionization corrections as predicted by the 
Cloudy models are generally small and within the typical error bars ($\sim$ 0.15 dex), in general agreement with previous 
studies. The Al III to Al II ratio for these and other absorbers from the literature are
compared, and we find that while the sub-DLAs have a larger scatter in the Al III
to Al II ratios than the DLAs, there appears to be little correlation between
the ratio and N$_{\rm H \ I}$. The relationship between the metallicity and the velocity width of the profile for these systems is 
investigated. We show that the sub-DLAs that have been observed to date follow a similar trend as DLA absorbers, with the more metal rich
systems exhibiting  large velocity widths. We also find that the systems at the upper edge of this relationship with high 
metallicities and large velocity widths are more likely to be sub-DLAs than DLA absorbers, perhaps implying that the sub-DLA
absorbers are more representative of massive galaxies. 

\end{abstract}

\begin{keywords}
{Quasars:} absorption lines-{ISM:} abundances
\end{keywords}

\section{Introduction}
The chemical composition of galaxies provides important clues into galaxy formation and evolution. Absorption systems 
in QSO spectra provide an opportunity to study the interstellar medium (ISM) and therefore the chemical enrichment of galaxies.
Damped Lyman-$\alpha$ (DLA) and sub-Damped Lyman-$\alpha$ (sub-DLA) absorbers are 2 classes of absorbers seen in 
the spectra of background QSOs. 
These high neutral hydrogen column density absorbers, particularly 
DLAs  with log N$_{\rm H \ I}$ $\ge$ 20.3 and sub-DLAs with 19 $\la$ 
log N$_{\rm H \ I}$ $<$ 20.3 \citep{Per01} are expected to contain the majority of the neutral gas in the universe, while the majority of
the baryons are thought to lie in the highly ionized and diffuse Lyman-$\alpha$ forest clouds with log N$_{\rm H \ I}$ $\la$ 14 \citep{Petit93}. 
The DLA and sub-DLA systems are believed to be strongly related to galaxies at any 
redshift. Indeed, several DLA host galaxies have been confirmed through deep imaging at $z \la 0.5$ eg; \citep{Chen03, Ghar06}. 

DLA systems have long been the preferred class of absorbers for chemical abundance studies. Observations by \citet{Des03}
of 12 sub-DLA absorbers at $1.8 < z < 4.3$ showed that these systems also contain metal line absorption features.  
Based on Fe II abundance measurements, \citet{Per03a} suggested stronger metallicity evolution in sub-DLA systems than DLA systems. 
To date, very few observations have been made of $z < 1.5$ 
sub-DLAs. Redshifts $z < 1.5$ span ~70$\%$ of the age of the universe (assuming a concordance cosmology of $\Omega_{m}=0.3, \Omega_{\Lambda}=0.7$). Clearly 
this redshift range is important to understanding the nature of sub-DLA systems and galactic chemical evolution as well.

Ionization effects can alter the abundances inferred for QSO absorbers. For DLAs, it is typically assumed that the ionization corrections
are small due to their large H I column densities that shield the cloud from ionizing photons. 
\citet{Des03} showed that the ionization corrections for sub-DLA systems are also typically small, with $< 0.2$ dex corrections to most ions. 
It was also shown by \citet{Per03b} that the contributions from DLAs to the neutral gas budget of the universe at $z > 3.5$ was lower
than at $z < 3.5$ and that sub-DLAs harbored this remaining HI. Recently \citet{Per06a, York06, Kul07, Pro06, Kh06} have suggested that sub-DLAs may be 
more metal rich than DLAs, and may contribute significantly to the cosmic metal budget. Sub-DLAs have thus become a topic of much interest.   

Many elements have been detected in QSO absorber systems, including C, N, O, Mg, Si, S, Ca, Ti, Cr, Mn, Fe, Ni, and Zn. Zn is 
often the preferred tracer of the gas-phase metallicity as it is relatively undepleted in the Galactic ISM, especially 
when the fraction of H in molecular form is low which is true in most DLAs. Zn also tracks the Fe abundance in Galactic stars, 
and the lines of Zn II $\lambda$$\lambda$ 2026,2062 are relatively weak and typically unsaturated.
These lines can also be covered with ground based spectroscopy over a wide range of redshifts, from 0.65 $\la$ z $\la$ 3.5, which covers a large portion of the 
history of the universe. Relative abundances of refractory elements such as Cr and Fe relative to Zn also relate to the amount of extinction. 

As galaxies mature, the ongoing processes associated with star formation such as stellar winds, supernovae explosions etc. 
expel metals into the interstellar medium. Over time, the interstellar metallicity is expected to reach a near solar value. Several models based on 
both analytical and semi-analytical calculations, as well as hydrodynamical simulations predict this increase to approximately solar values. 
Recent findings indicate that the N$_{\rm H \ I}$ weighted mean metallicity (Z) of DLAs is well below solar ($\sim$ 0.1Z$_{\sun}$ 
at $z=2.0$ and $\sim$0.16Z$_{\sun}$ at $z=$1) and shows little if any evolution that is predicted by models \citep{Kul05}.
Some lines of investigation are now suggesting that sub-DLAs may contain a significant portion of these 
``missing metals'', and may be more representative of typical massive spirals or ellipticals \citep{Kul07, Kh06}. \citet{Wol93} claimed 
that the DLAs can be considered as progenitors to normal modern spirals, and argued that the line 
profiles suggest rotation curves (in absorption). Others however note that the absence of large galaxies in a number of
searches for DLA host galaxies, suggesting that many are dwarf galaxies often with low impact parameters and lost in the light of the QSO \citep{York86, Hop05}.
\citet{Kul07} show that based on available spectroscopic measurements, at $z \la 1$ the contribution of sub-DLAs to the total metal budget may be several times 
greater than that of DLAs. For a constant dust-to-gas ratio, the dust obscuration bias is also expected to be less for sub-DLAs than 
for DLAs, therefore one might expect to be able to find metal rich sub-DLAs more frequently if dust obscuration effects are important.

Although difficult to constrain, it is generally acknowledged that there may be a significant ionization fraction in the sub-DLA systems, 
especially for systems with lower N$_{\rm H \ I}$ \citep{Per03a, Omear06}. It has also been claimed that the sub-DLAs contain $\sim$15$\%$ of the H atoms in
the universe, and that they may be a distinct population of absorbers separated from the DLA absorbers \citep{Omear06}.
The H I column density distribution for QSO absorbers shows that systems with  
19 $\la$ log N$_{\rm H \ I} \la$ 20.3 are $\sim$8 times more numerous than the higher column density classical damped Ly$\alpha$ systems, 
making sub-DLAS more readily available probes of neutral gas in the distant Universe.

In this paper we present high-resolution measurements of 7 sub-DLAs and 1 DLA taken with the Magellan Inamori Kyocera Echelle (MIKE) 
spectrograph  on the 6.5m Clay telescope at Las Campanas observatory. The structure of this paper is as follows. 
In $\S$ 2, we discuss details of our observations and data reduction techniques. In $\S$ 3, column density determinations are discussed. $\S$ 4 gives information 
on the individual objects, while in $\S$ 5 we discuss the abundances of these absorbers, the Al III to Al II ratio, and ionization corrections,
in $\S$ 6 we draw some general conclusions.

\section{Observations and Data Reduction}

The spectra presented here were obtained over 2 separate epochs, 2005 Sep. and 2006 February, respectively, with the 6.5m Magellan 
Clay telescope at Las Campanas Observatory. These objects were observed with the Magellan Inamori Kyocera Echelle 
spectrograph (MIKE) \citep{Bern03}. This is a double sided spectrograph with both a blue and a red camera, providing for 
simultaneous wavelength coverage from $\sim$3340 \ang to $\sim$9400 \ang. Targets were observed in 
 multiple exposures of 1800 to 2700 sec to minimize cosmic ray defects. The seeing was typically 
 $<$ 1$\arcsec$, averaging $\sim$ 0.7$\arcsec$. All of the target QSOs were observed 
 with the 1$\arcsec$x5$\arcsec$ slit and the spectra were binned 2x3 (spatial by spectral) during readout. 
The resolving power of the MIKE spectrograph is $\sim$19,000 and $\sim$25,000 on the red and blue sides respectively with a 1$\arcsec$x5$\arcsec$ slit.
Table 1 gives a summary of the observations. 

These spectra were reduced using the MIKE pipeline reduction code in IDL developed by S. Burles, J. X. Prochaska, 
and R. Bernstein. Wavelengths were calibrated using a Th-Ar comparison lamp taken after each exposure. 
The MIKE software first bias subtracts from the overscan region and flat-fields the data, then sky-subtracts and extracts the spectral orders 
using the traces from flat field images. The pipeline calibration software also corrects for heliocentric velocities and converts 
the wavelengths to vacuum values. Each individual order was then extracted from the IDL structure created by the pipeline software and 
combined in IRAF using rejection parameters to reduce the effects of cosmic rays. These combined spectra were then normalized using a
Legendre polynomial function to fit the continuum. Typically, these functions were of order five or less. 

Our sample consists of 7 sub-DLAs and 1 DLA at z$<$1.5, where there are few existing abundance measurements, 
including 3 absorbers with z$<$1.0 where there are even fewer abundance measurements 
\citep{Kh04, Per06a, Per06b, Pet00}. The objects from our sample were chosen partly because of strong lines of Mg 
II $\lambda\lambda$ 2796, 2803, or Fe II $\lambda$ 2600 seen in SDSS spectra, 
or otherwise previously measured and given in \citet{Rao06}. 
These sub-DLAs and the two observed in \citet{Per06a, Per06b} comprise all of high resolution observations of 
sub-DLAs at $z < 1.5$ to date. Four of the target QSOs in this sample were observed in the Sloan Digital Sky Survey
(SDSS) and all have a known N$_{\rm H \ I}$ from HST spectra. H I column densities were taken from \citet{Rao06} and references therein.
We provide an appendix at the end of this paper showing the fits of the Lyman-$\alpha$ profiles from the UV spectra. 
Throughout this paper, the QSO names are given in J2000 coordinates. 

\begin{table*}
\begin{minipage}{170mm}
\begin{center}	
{\scriptsize
\caption{Summary of Observations}
\begin{tabular}{ccccccccccc}
\hline
\hline
QSO		&	Original or SDSS ID		&	RA		&	Dec		&	m$_{V}$	&$z_{em}$	&$z_{abs}$	&	N$_{\rm H \ I}$ &	Exposure Time	&	Epoch			& Mg II Reference		\\
J2000		&					&			&			&		&		&		&	cm$^{-2}$ 	&	Sec		&				&		\\		
\hline
Q0354-2724	&	Q0352-272			&	03:54:05.9	&	-27:24:25.7	&	17.9	&	2.823	&	1.4051	&	20.18$\pm$0.15	&	10800		&	2005 Sep, 2006 Feb.	& 1		\\
Q0826-2230	&	Q0823-223			&	08:26:01.5	&	-22:30:26.2	&	16.2	&	$>$0.911&	0.9110	&	19.04$\pm$0.04	&	8627		&	2006 Feb.		& 2 		\\
Q1009-0026	&	SDSS J100930.46-002619.1	&	10:09:30.4	&	-00:26:19.1	&	17.4	&	1.244	&	0.8426	&	20.20$\pm$0.06	&	8100		&	2006 Feb.		& 3		\\
$\cdots$	&	$\cdots$			&	$\cdots$	&	$\cdots$	&	$\cdots$&	$\cdots$&	0.8866	&	19.48$\pm$0.05	&	$\cdots$	&	$\cdots$		& 3		\\
Q1010-0047	&	SDSS J101033.44-004724.5	&	10:10:33.4	&	-00:47:24.5	&	18.0	&	1.671	&	1.3270	&	19.81$\pm$0.05	&	7191		&	2006 Feb.		& 3		\\
Q1224+0037	&	SDSS J122414.29+003709.0	&	12:24:14.3	&	00:37:09.0	&	18.7	&	1.482	&	1.2346	&	20.88$\pm$0.05	&	5400		&	2006 Feb.		& 3		\\
$\cdots$ 	&	$\cdots$			&	$\cdots$	&	$\cdots$	&	$\cdots$&	$\cdots$&	1.2665	&	20.00$\pm$0.07	&	$\cdots$	&	$\cdots$		& 3		\\
Q2331+0038	&	SDSS J233121.81+003807.4	&	23:31:21.8	&	00:38:07.4	&	17.8	&	1.486	&	1.1414	&	20.00$\pm$0.05	&	8000		&	2005 Sep 		& 3		\\
\hline
\end{tabular}

}
\end{center}
Notes -- Emission redshifts are from the reference below, or if the target was observed in the SDSS, the emission redshift is from \citet{Schneid05}\\
Mg II References. -- (1) \citet{Sarg89}, (2) \citet{Fal90}, (3) \citet{Nest04}

\end{minipage}
\end{table*}

\begin{table*}
\begin{minipage}{170mm}
\begin{center}
{\scriptsize
\caption{Rest-frame equivalent width measurements of key metal lines. Values and 1$\sigma$ errors are in units of m\ang.}
\begin{tabular}{lcccccccccccccc}
\hline
\hline
QSO	&	z$_{abs}$	&	Mg I		 	&	Mg II		 	&	Mg II 		 	&	Al II 		 	&	Al III 		 	&	Al III 		 	&	Si II 		 	&	Si II 		 	&	Ca II 		 	&	Ca II			&	Cr II 			\\
\hline
	&			&	2852			&	2796			&	2803			&	1670			&	1854			&	1862			&	1526			&	1808			&	3933			&	3969			&	2056		 	\\
\hline
Q0354-2724	&	1.4051	&	537	$\pm$	9	&	2665	$\pm$	14	&	2394	$\pm$	10	&	$<1114^{a}$	 	&	$<412^{a}$	 	&	$<607^{a}$	 	&	$<1670^{a}$	 	&	$\cdots$		&	$\cdots$	 	&	$\cdots$	 	&	64	$\pm$	8	\\
Q0826-2230	&	0.9110	&	123	$\pm$	9	&	1280	$\pm$	16	&	892	$\pm$	15	&	$\cdots$	 	&	$<$9	 	 	&	$<$9	 	 	&	$\cdots$	 	&	$<$10		 	&	22	$\pm$	4	&	14	$\pm$	4	&	$<$5	 	 	\\
Q1009-0026	&	0.8426	&	77	$\pm$	14	&	713	$\pm$	13	&	551	$\pm$	13	&	$\cdots$	 	&	77	$\pm$	8	&	33	$\pm$	6	&	$\cdots$	 	&	$\cdots$		&	$\cdots$	 	&	$\cdots$	 	&	$<$10	 	 	\\
Q1009-0026	&	0.8866	&	308	$\pm$	22	&	1792	$\pm$	16	&	1508	$\pm$	18	&	$\cdots$	 	&	146	$\pm$	10	&	88	$\pm$	13	&	$\cdots$	 	&	$<$12		 	&	103	$\pm$	16	&	83	$\pm$	14	&	$<$5	 	 	\\
Q1010-0047	&	1.3270	&	343	$\pm$	18	&	2040	$\pm$	14	&	1768	$\pm$	15	&	883	$\pm$	17	&	279	$\pm$	9	&	141	$\pm$	8	&	681	$\pm$	21	&	44	$\pm$	11	&	$\cdots$	 	&	 $\cdots$ 	 	&	$<$9	 	 	\\
Q1224+0037	&	1.2346	&	212	$\pm$	15	&	1086	$\pm$	15	&	1032	$\pm$	18	&	513	$\pm$	17	&	159	$\pm$	10	&	99	$\pm$	9	&	477	$\pm$	22	&	52	$\pm$	7	&	$\cdots$	 	&	$\cdots$ 	 	&	47	$\pm$	7	\\
Q1224+0037	&	1.2665	&	188	$\pm$	47	&	2181	$\pm$	28	&	1876	$\pm$	25	&	827	$\pm$	15	&	84	$\pm$	24	&	$<$10	 	 	&	$<$1140$^{b}$ 		&	$<12$		 	&	$\cdots$	 	&	 $\cdots$ 	 	&	$<$9	 	 	\\
Q2331+0038	&	1.1414	&	299	$\pm$	19	&	2534	$\pm$	43	&	2066	$\pm$	47	&	424	$\pm$	85	&	105	$\pm$	50	&	52	$\pm$	25	&	$\cdots$	 	&	$<$13		 	&	257	$\pm$	23	&	141	$\pm$	26	&	$<$9	 	 	\\
\hline																																															
\hline																																															
QSO	&	z$_{abs}$	&	Mn II 		 	&	Mn II 		 	&	Mn II 		 	&	Fe II 		 	&	Fe II 		 	&	Fe II		 	&	Fe II 		 	&	Fe II 		 	&	Fe II 		 	&	Zn II$^{c}$ 		&	Zn II$^{d}$		\\
\hline
	&			&	2576			&	2594			&	2606			&	2260			&	2344			&	2374			&	2382			&	2586			&	2600			&	2026			&	2062			\\
\hline
Q0354-2724	&	1.4051	&	126	$\pm$	13	&	100	$\pm$	10	&	75	$\pm$	8	&	109	$\pm$	9	&	1249	$\pm$	18	&	739	$\pm$	17	&	1657	$\pm$	21	&	1143	$\pm$	16	&	1715	$\pm$	18	&	84	$\pm$	10	&	68	$\pm$	15	\\
Q0826-2230	&	0.9110	&	$<$5	 		&	$<$5	 	 	&	$<$5	 	 	&	$<$4	 	 	&	161	$\pm$	21	&	39	$\pm$	15	&	444	$\pm$	14	&	102	$\pm$	10	&	353	$\pm$	20	&	44	$\pm$	8	&	26	$\pm$	14	\\
Q1009-0026	&	0.8426	&	27	$\pm$	4	&	42	$\pm$	6	&	$<$8	  	 	&	31	$\pm$	6	&	275	$\pm$	6	&	179	$\pm$	5	&	389	$\pm$	4	&	265	$\pm$	6	&	403	$\pm$	5	&	$<$13	 	 	&	$<$10	 	 	\\
Q1009-0026	&	0.8866	&	$<$5	 		&	$<$5	 	 	&	$<$5	 	 	&	23	$\pm$	4	&	625	$\pm$	10	&	255	$\pm$	10	&	1000	$\pm$	12	&	568	$\pm$	11	&	981	$\pm$	18	&	41	$\pm$	6	&	$<$5	 	 	\\
Q1010-0047	&	1.3270	&	$<$15	 		&	$<$15	  	 	&	$<$15	 	 	&	0	 	 	&	905	$\pm$	32	&	396	$\pm$	24	&	1230	$\pm$	14	&	741	$\pm$	17	&	1269	$\pm$	13	&	$<$9	 	 	&	$<$6	 	  	\\
Q1224+0037	&	1.2346	&	$<$29	 		&	$<$29	 	 	&	$<$29	 	 	&	$<$40	 	 	&	632	$\pm$	34	&	429	$\pm$	35	&	758	$\pm$	24	&	658	$\pm$	29	&	827	$\pm$	29	&	$<$14	 	 	&	30	$\pm$	7	\\
Q1224+0037	&	1.2665	&	$<$31	 		&	$<$31	 	 	&	$<$31	 	 	&	$<$30	 	 	&	757	$\pm$	28	&	271	$\pm$	40	&	1206	$\pm$	36	&	736	$\pm$	65	&	1385	$\pm$	27	&	$<$13	 	 	&	$<$9	 	 	\\
Q2331+0038	&	1.1414	&	$<$13	 		&	$<$13	 	 	&	$<$13	 	 	&	$<12$	 	 	&	509	$\pm$	13	&	324	$\pm$	17	&	1206	$\pm$	46	&	662	$\pm$	36	&	1285	$\pm$	36	&	29	$\pm$	10	&	$<$7	 	 	\\

\hline
\end{tabular}
}
\end{center}
$^{a}$Lines were blended with Lyman-$\alpha$ forest lines.\\ 
$^{b}$Blended with another feature.\\
$^{c}$This line is a blend with Mg I$\lambda$ 2026, although the Mg I contribution is judged to be insignificant in all cases.\\
$^{d}$As this line is blended with the Cr II $\lambda$ 2062 line, this value represents the total EW of the line.\\
\end{minipage}
\end{table*}

\section{Determination of Column Densities}
    Column densities were determined from profile fitting with the package FITS6P \citep{Wel91}, which has
evolved from the code by Vidal-Madjar et al (1977). FITS6P iteratively minimizes the $\chi^{2}$ value between the data and
a theoretical Voigt profile that is convolved with the instrumental profile. 
The profile fit used multiple components, tailored to the individual system. For the central, core components, the Doppler parameters
($b_{eff}$) and radial velocities were determined from the weak and unsaturated lines, typically the Mg I $\lambda$ 2852 line. 
For the weaker components at higher radial velocities the $b_{eff}$ and component velocity values were determined from stronger transitions such as the 
Fe II $\lambda\lambda$ 2344, 2382 lines and the Mg II $\lambda\lambda$ 2796, 2803 lines. A set of $b_{eff}$ and $v$ values were thus
determined that reasonably fit all of the lines observed in the system. 
The atomic data used in line identification and profile fitting are from \citet{Morton03}.

	 In general, if a multiplet was observed, the lines were fit simultaneously until convergence. 
For all of the systems, the Fe II $\lambda$ 2344, 2374, 2382 lines were fit simultaneously to determine a set of column
densities that fit the spectra reasonably well. Similarly, the Mg II $\lambda\lambda$ 2796, 2803 lines were also fit together. 
The Zn II $\lambda$ 2026 line is blended with a line of Mg I. The Mg I contribution to the line was estimated using the
Mg I $\lambda$ 2852 line, for which f$\lambda$ $\sim$32 times that of the $\lambda$ Mg I 2026 line. 
The Zn II components were then allowed to vary while the Mg I components were held fixed. 
N$_{\rm Cr \ II}$ was determined by simultaneously fitting the Cr II $\lambda$ 2056 line and the blended Cr II + Zn II $\lambda$ 2062 line, where the 
contribution from Zn II was estimated from the Zn II + Mg I $\lambda$ 2026 line. See also \citet{Kh04} for a discussion of the profile fitting scheme.
In this paper we adopt the standard notation 
	$$[X/Y] = log (N_{\rm X}/N_{\rm H \ I}) - log (X/H)_{\sun}.$$
	Solar system abundances have been adopted from \citet{Lodd03}.

   Using the package SPECP, also developed by D.E. Welty, we determined equivalent widths as well as column densities via the apparent 
optical depth method (AOD) \citep{Sav96}. We present the rest-frame equivalent widths ($W_{0}$) of the lines in Table 4. 
The 1$\sigma$ errors for the equivalent widths are given also and reflect both uncertainties in the 
continuum level and in the photon noise. If a certain line was not detected, the limiting equivalent width was determined from 
the local signal to noise ratio (S/N). Assuming a linear curve of growth, a 3$\sigma$ upper limit was also placed on the column density.

\section{Discussion of Individual Objects}
\subsection{Q0354-2724 ($z_{em}$ = 2.823)}
   This QSO (M$_{V}$=17.9) has a known sub-DLA with log N$_{\rm H \ I}$=20.18 at z$_{abs}$=1.4051 \citep{Sarg89, Rao06}. We observed this target 
during both the 2005 Sep. and 2006 Feb. epochs for a total exposure time of 10800 sec. We detect strong absorption features of Mg I, 
Mg II, Cr II, Mn II, Fe II, and Zn II. The S/N varies from $\sim$20 at $\sim$ 4900 \ang to $\sim$50 at $\sim$5700 \ang. 
Several lines of interest such as Al II $\lambda$ 1670, Al III $\lambda\lambda$ 1854, 1862
and Si II $\lambda$ 1808 were inside the Lyman-$\alpha$ forest due to the relatively high redshift of the background QSO. 
Figure 1 shows velocity plots of several lines of interest for this system. 
There are no detectable components from the absorption system in any species outside the plotted range. 
Sargent et al. (1989) reported the existense of a C IV absorption
system at $z_{abs}=2.1442$ based on low-resolution ($\sim$ 6 \ang) observations. Based on these higher resolution spectra, we believe that this system was 
misidentified. There is a small feature at $\sim$350 \kms  of the Zn II $\lambda$ 2026 line ($\lambda_{obs}$= 4878.7 \ang), as can be seen in figure 1, with W$_{\rm obs}$=67 m\ang.
However, if this line is assumed to be the 
C IV $\lambda$ 1550 line, then the $\lambda$ 1548 line should be $\sim$4 \ang blueward of the Zn II 2026 line. At the resolution of Sargent et al. (1989), 
the Zn II line from the $z_{abs}=1.4051$ sub-DLA was likely confused for this C IV $\lambda$ 1548 line. This feature is unidentified, it does not match  
any possible lines from the other known systems.

  This system shows a complex velocity structure with a total of at least 8 components necessary to fit the observed profile.
The total profile spans $\sim$360 \kms (see $\S$ 5.4 for a more detailed discussion of how we derived the velocity width).  We detect Zn II at S/N $\sim$ 20 with log
N$_{\rm Zn \ II}$=12.73$\pm$0.03 for this sub-DLA giving a nearly solar abundance [Zn/H]=-0.08. 
The Mg I contribution to the blended Zn II + Mg I $\lambda$ 2026 line was estimated from the Mg I $\lambda$ 2852 line. 
In theory, if the Mg I $\lambda$ 2852 line is saturated, which is rare but does occur occasionally for DLAs, the 
estimated contribution to the blended Zn II + Mg I $\lambda$ 2026 line from the Mg I $\lambda$ 2852 line could be lower than its true value. 
The EW of the Mg I $\lambda$ 2852 line was W$_{rest}$ = 537 m\ang, which is below the 
600 m\ang limit where the line begins to becomes significantly saturated \citep{York06}. 
This system also shows moderate iron depletion with [Fe/Zn]=-0.43. Table 3 gives the results of the profile fitting analysis for each component. 

\begin{figure}
\includegraphics[height=\linewidth, angle=90]{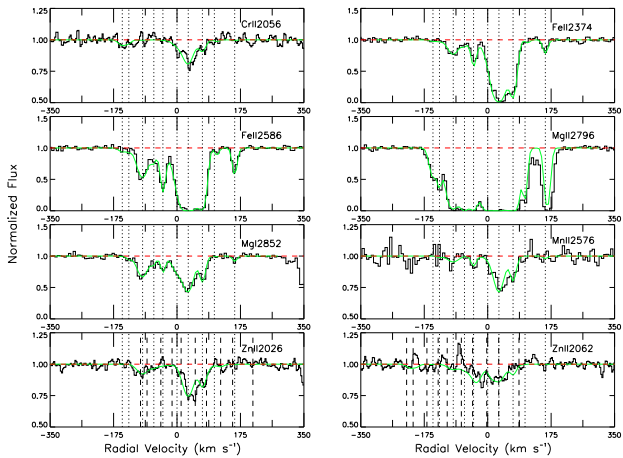}
\caption{Velocity plots for several lines of interest in the $z=$1.4051 system in the spectrum of Q0354-2724. The
 solid green line indicates the theoretical profile fit to the spectrum, and the dashed red line is the continuum level.
 The vertical dotted lines indicate the positions of the components that were used in the fit. In the case of the Zn II $\lambda\lambda$ 2026,2062
 lines, the dashed vertical lines indicate the expected positions of the components of the Mg I ($\lambda$ 2026) and Cr II ($\lambda$ 2062) contributions 
 respectively.}
\end{figure}

\begin{table*}
\begin{minipage}{135mm}
\begin{center}
\caption{Column Densities for the $z=$1.4051 absorber in Q0354-2724. Velocities and b$_{eff}$ values are in units of \kms, and column
  	 densities are in units of cm$^{-2}$.}
\begin{tabular}{lccccccc}
\hline
\hline
Vel	&	b$_{eff}$	&	Mg I	&	Mg II	&	Cr II	&	Mn II	&	Fe II	&	Zn II	\\
\hline
-151	&	10.9	&	-			&	(3.32$\pm$0.53)E+12	&	(2.07$\pm$0.68)E+12	&	-			&	(3.24$\pm$0.90)E+12	&	-			\\
-133	&	5.0	&	-			&	(3.22$\pm$0.77)E+12	&	-			&	-			&	-			&	-			\\
-95	&	18.7	&	(1.04$\pm$0.06)E+12	&	$>$2.85E+13		&	-			&	(7.09$\pm$2.09)E+11	&	(4.89$\pm$0.20)E+13	&	(7.61$\pm$1.29)E+11	\\
-64	&	6.4	&	(1.63$\pm$0.38)E+11	&	$>$1.57E+14		&	-			&	-			&	(9.20$\pm$1.10)E+12	&	-			\\
-39	&	6.3	&	(3.67$\pm$0.43)E+11	&	$>$1.21E+14		&	-			&	(6.10$\pm$1.53)E+11	&	(5.66$\pm$0.39)E+13	&	-			\\
0	&	17.4	&	(6.64$\pm$0.63)E+11	&	$>$4.71E+13		&	(2.73$\pm$0.79)E+12	&	-			&	(4.62$\pm$0.31)E+13	&	-			\\
31	&	18.2	&	(2.10$\pm$0.09)E+12	&	$>$6.80E+13		&	(1.37$\pm$0.09)E+13	&	(4.80$\pm$0.27)E+12	&	(8.53$\pm$0.87)E+14	&	(3.85$\pm$0.16)E+12	\\
70	&	8.7	&	(7.76$\pm$0.06)E+11	&	$>$6.78E+14		&	(3.86$\pm$0.64)E+12	&	(1.85$\pm$0.19)E+12	&	(3.72$\pm$0.64)E+14	&	(8.16$\pm$1.09)E+11	\\
103	&	3.1	&	-			&	$>$8.11E+13		&	-			&	-			&	(4.13$\pm$0.89)E+12	&	-			\\
159	&	4.9	&	(9.96$\pm$3.31)E+10	&	$>$6.64E+13		&	-			&	-			&	(2.25$\pm$0.19)E+13	&	-			\\

\hline
\end{tabular}
\end{center}
\end{minipage}
\end{table*}

\subsection{Q0826-2230 ($z_{em}$ $>$ 0.911)}
This is a BL Lac object. Due to the ambiguous emission redshift for this object, it is possible that this sub-DLA system is associated with the source itself. 
This object has a weak sub-DLA with log N$_{\rm H \ I}$=19.04 at $z=$0.9110 \citep{Fal90, Rao06}. A total of 7 components were used in the theoretical profile. 
The total velocity width of the profile was $\sim$322 \kms.
Figure 2 shows velocity plots for some of the lines of interest. We detect strong features of Mg II, Fe II, and Ca II. Mg I is also 
unambiguously detected, but the majority of the profile is contained in the component at $v = -65$ \kms. We also detect Zn II
$\lambda$ 2026 and 2062 at S/N $\sim$ 40 in the region, with the majority of the contribution from the $v = -65$ \kms component.
There was a null detection of the Cr II $\lambda$ 2056 line, 
which is the intrinsically strongest line available, so only a limit could be placed on N$_{\rm Cr  \ II}$. 

This system has super-solar metallicity with [Zn/H]=+0.68 (see the appendix for a discussion of the H I values). 
The weaker Fe II $\lambda\lambda$ 2249, 2260 lines were
not detected in this system, but the Fe II $\lambda$ 2374 line was unsaturated (W$_{rest}$=39 m\ang, spread over 1 \ang) and an accurate column density 
could be determined. This system appears to have a high depletion with [Fe/Zn]=-1.63. The measured Ca II column density of log N$_{\rm Ca \ II}$=11.75
likely underestimates the true Ca column density because the ionization potential for
Ca II is 11.868 eV is less than the ionization potential of H I. Table 4 gives the results of the Voigt profile fitting for each component. 

Grids of Cloudy models \citep{Fer98} were constructed for this system which were tailored to fit the N$_{\rm H \ I}$ and metallicity observed. 
The Al II $\lambda$ 1670 line for this system was not in the wavelength section covered, and only limits could be placed on the 
Al III $\lambda\lambda$ 1854, 1862 lines. We were however able to constrain
the ionization parameter U based on the Mg II to Mg I ratio observed. Since the Mg II $\lambda\lambda$ lines were saturated, 
this ratio is only a lower limit, and the subsequent ionization corrections may be different. However, the Mg II lines were not optically thick 
at this resolution, and the true column density may not be much higher than the value given. The AOD column density also agrees very well 
with the profile fitting column density. Based on the model predictions, the ionization parameter is small and the 
ionization corrections for Zn are negligible ($<$0.1 dex) for a large range of ionization parameters.
See $\S$5.3 for a more thorough account of the Cloudy modeling and ionization corrections.

\begin{figure}
\includegraphics[height=\linewidth,  angle=90]{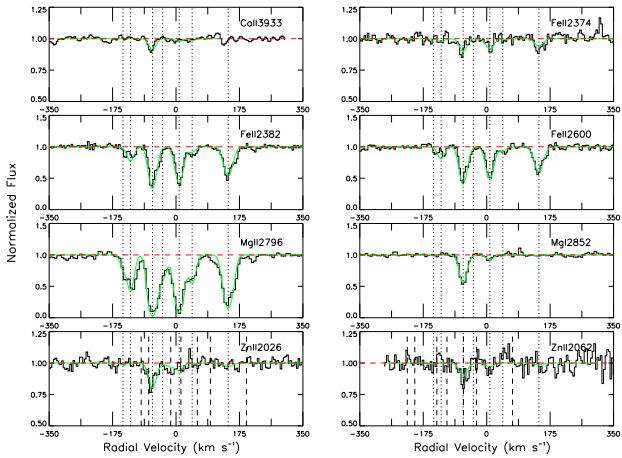}
\caption{The same as fig.1, but for the z=0.9110 system in the spectrum of Q0826-2230}
\end{figure}

\begin{table*}
\begin{minipage}{115mm}
\begin{center}
\caption{Same as table 3, but for the z=0.9110 absorber in Q0826-2230.}
\begin{tabular}{ccccccc}
\hline
\hline
Vel	&b$_{eff}$	&	Mg I 			&	Mg II 			&	Ca II 			&	Fe II 			&	Zn II	\\
\hline
-146	&	9.4	&	 - 			&	(9.77$\pm$1.72)E+11	&	-			&	-			&	-	\\
-126	&	11.6	&	-			&	(4.22$\pm$0.28)E+12	&	-			&	(2.94$\pm$0.74)E+12	&	-	\\
-65	&	13.1	&	(1.08$\pm$0.05)E+12	&	$>$1.94E+13		&	(3.37$\pm$0.19)E+11	&	(1.37$\pm$0.10)E+13	&	(1.82$\pm$0.19)E+12\\
-37	&	14.7	&	-			&	(2.86$\pm$0.26)E+12	&	-			&	-			&	-	\\
9	&	13.8	&	(1.42$\pm$0.30)E+11	&	$>$1.51E+13		&	(8.63$\pm$1.83)E+10	&	(1.14$\pm$0.09)E+13	&	(4.26$\pm$1.74)E+11\\
45	&	12.0	&	-			&	(3.07$\pm$0.23)E+12	&	(5.38$\pm$1.71)E+10	&	-			&	-	\\
144	&	15.4	&	-			&	$>$1.21E+13		&	(8.00$\pm$1.92)E+10	&	(8.98$\pm$0.90)E+12	&	-	\\

\hline
\end{tabular}
\end{center}
\end{minipage}
\end{table*}

\subsection{Q1009-0026 (z$_{em}$ = 1.244)}
 This system has two confirmed sub-DLAs with log N$_{\rm H \ I}$=20.20 at $z=0.8426$ (system A) and log N$_{\rm H \ I}$=19.48 at 
$z=0.8866$ (system B) \citep{Nest04, Rao06}. System A shows a relatively simple velocity structure with only 4 components needed to fit the observed profile. 
The profile for system A spans a total of $\sim$94 \kms. 
For this system, the Al  II line was below the wavelength region covered by the spectrograph, but the Al III 
$\lambda\lambda$ 1854, 1862 lines were observed although the dominant species is likely Al II. This system also shows
fairly strong Mn II $\lambda\lambda$ 2576, 2594, 2606 lines. 
We did not detect any Zn II $\lambda\lambda$ 2026 lines with S/N $\sim$ 20 in the region, so we give the metallicity as an upper limit. 
This system has a low metallicity, with [Zn/H]$<$-0.98, [Fe/H]=-1.28, and [Mn/H]=-1.44 although both Mn and Fe are often depleted in the Galaxy.

System B shows a more complicated velocity structure with 7 components needed to fit the profile. 
This system has a larger velocity width, spanning $\sim$334 \kms.
Again, the Al II $\lambda$ 1670 line was not in the wavelength region, but Al III $\lambda\lambda$ 1854, 1862 lines were detected.
A S/N $\sim$ 25 was reached in the region of Zn II $\lambda$ 2026 which was detected at $\sim$ 5$\sigma$ with an equivalent width W$_{rest}=41$ m\ang.
Profile fitting of the line gives log N$_{\rm Zn \ II}=12.36$ and [Zn/H]=+0.25. 
We were able to constrain the ionization parameter for this system to be log U=-3.70 based on the observed Al III to Fe II ratio (see $\S$ 5.3 for a more 
detailed discussion of the method).  
The Ca II $\lambda\lambda$ 3933, 3969 lines were also detected in this system with log N$_{\rm Ca \ II}$=12.26. Mn was not detected in this 
system with [Mn/Fe]$<$-1.24. Tables 5 and 6 show the results of the Voigt profile decomposition for systems A and B respectively.

\begin{figure}
\includegraphics[height=\linewidth, angle=90]{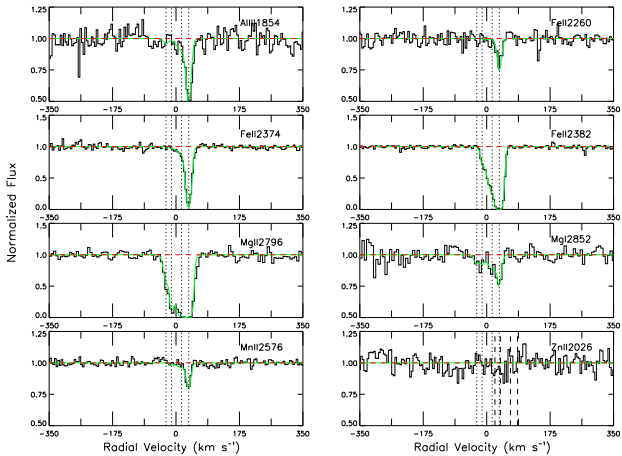}
\caption{The same as fig.1, but for the z=0.8426 system in the spectrum of Q1009-0026}
\end{figure}

\begin{table*}
\begin{minipage}{115mm}
\begin{center}
\caption{Same as table 3, but for the z=0.8426 absorber in Q1009-0026.}
\begin{tabular}{lcccccc}
\hline
\hline
	Vel	&b$_{eff}$	&	Mg I 			&	Mg II			&	Al III			&	Mn II			&	Fe II			\\
\hline
	-28	&	8.7	&	-			&	(1.95$\pm$0.21)E+12	&	-			&	-			&	- 			\\
	-13	&	5.6	&	-			&	(5.74$\pm$0.66)E+12	&	-			&	-			&	(1.53$\pm$0.32)E+12	\\
	15	&	16.7	&	(2.60$\pm$0.64)E+11	&	$>$2.83E+13		&	(1.16$\pm$0.28)E+12	&	(4.89$\pm$1.11)E+11	&	- 			\\
	35	&	7.2	&	(2.73$\pm$0.50)E+11	&	$>$1.65e+14		&	(4.20$\pm$0.33)E+12	&	(1.34$\pm$0.08)E+12	&	(2.47$\pm$0.23)E+14	\\

\hline
\end{tabular}
\end{center}
\end{minipage}
\end{table*}

\begin{table*}
\begin{minipage}{145mm}
\begin{center}
\caption{Same as table 3, but for the z=0.8866 absorber in Q1009-0026.}
\begin{tabular}{lcccccccc}
\hline
\hline
	Vel	&	b$_{eff}$&	Mg I 			&	Mg II			&	Al III			&	Ca II			&	Fe II		&	Zn II				\\
\hline
	-121	&	4.6	&	-			&	(4.47$\pm$0.49)E+12	&	-			&	-			&	(4.69$\pm$0.57)E+12	&	-			\\
	-66	&	15.6	&	(6.22$\pm$0.65)E+11	&	$>$7.77E+13		&	(2.51$\pm$0.24)E+12	&	(4.58$\pm$0.65)E+11	&	(3.22$\pm$0.11)E+13	&	-			\\
	-13	&	23.9	&	(1.72$\pm$0.09)E+12	&	$>$2.02E+14		&	(9.99$\pm$0.36)E+12	&	(1.35$\pm$0.08)E+12	&	(3.31$\pm$0.31)E+14	&	(2.28$\pm$0.26)E+12	\\
	70	&	5.2	&	(1.99$\pm$0.48)E+11	&	$>$2.49E+13		&	(6.71$\pm$2.13)E+11	&	-			&	(2.40$\pm$0.52)E+12	&	-			\\
	87	&	3.6	&	-			&	$>$1.32E+13		&	-			&	-			&	(8.16$\pm$0.79)E+12	&	-			\\
	169	&	4.4	&	-			&	(1.99$\pm$0.31)E+12	&	-			&	-			&	(1.93$\pm$0.57)E+12	&	-			\\
	182	&	5.1	&	-			&	(9.47$\pm$1.23)E+12	&	-			&	-			&	-			&	-			\\

\hline
\end{tabular}
\end{center}
\end{minipage}
\end{table*}

\begin{figure}
\includegraphics[height=\linewidth, angle=90]{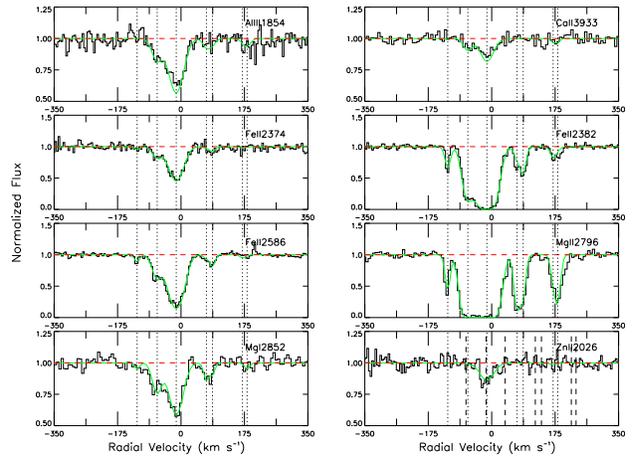}
\caption{The same as fig.1, but for the z=0.8866 system in the spectrum of Q1009-0026}
\end{figure}

\subsection{Q1010-0047 ($z_{em} = 1.671$)}
This system has a confirmed sub-DLA with log N$_{\rm H \ I}$=19.81 at $z=1.3270$ \citep{Nest04, Rao06}. This system was fit with a 8 component model.
The profile spanned $\sim$ 308 \kms in total.  
We detect strong lines of Mg II, Al II, Al III, and Fe II in this system but no detection
of Zn II $\lambda$ 2026 in the region with S/N $\sim$ 30. Table 7 shows the 
results of the profile fitting for each component. From the 3$\sigma$ limiting equivalent width of W$_{rest}=9$ m\ang, 
this system has log N$_{\rm Zn \ II}<$11.67 and [Zn/H]$<$-0.75. We detect strong Si II $\lambda\lambda$ 1526, 1808 lines with log N$_{\rm Si \ II}$=15.02. 
This system also shows apparent $\alpha$-enhancement with [Si/Fe]=0.42. Although depletion may be important, the metallicities based on
Si and Fe are [Si/H]=-0.33 and [Fe/H]=-0.75.
As there were detections of both Al II and Al III in this system, we were able to constrain the ionization parameter U for this system 
to be log U$\la$-4.15. See $\S$ 5.4 and $\S$ 5.5 for a more complete discussion of the Cloudy modeling and ionization corrections.   
Figure 5 gives velocity plots for some of the lines of interest in this system. The feature at $\sim$-300 \kms in the plot of Al III $\lambda$ 1862 
is unidentified. There are several other systems seen in the spectrum of this QSO, but no known lines match the 
observed wavelength of this feature.

\begin{table*}
\begin{minipage}{115mm}
\caption{Same as table 3, but for the $z=1.3270$ absorber in Q1010-0047.}
\begin{tabular}{lccccccc}
\hline
\hline
Vel	&	b$_{eff}$	&	Mg I			&	Mg II			&	Al II		&	Al III			&	Si II			&	Fe II			\\
\hline
-125	&	4.9		&	(1.53$\pm$0.46)E+11	&	(8.63$\pm$1.59)E+11	&	-		&	-			&	-			&	-			\\
-85	&	16.3		&	(7.44$\pm$0.73)E+11	&	$>$3.48E+13		&	$>$7.87E+12	&	(4.06$\pm$0.37)E+12	&	(2.82$\pm$0.47)E+14	&	(8.72$\pm$0.47)E+13	\\
-55	&	4.2		&	-			&	$>$4.74E+14		&	$>$5.45E+12	&	-			&	-			&	(6.21$\pm$1.61)E+12	\\
-35	&	10.2		&	-			&	$>$2.02E+13		&	$>$3.55E+12	&	(2.91$\pm$0.37)E+12	&	-			&	(1.65$\pm$0.18)E+13	\\
-14	&	3.9		&	(3.58$\pm$0.62)E+11	&	$>$4.21E+14		&	$>$4.96E+12	&	(2.97$\pm$0.48)E+12	&	-			&	(1.56$\pm$0.26)E+13	\\
4	&	7.0		&	-			&	$>$3.47E+14		&	$>$1.64E+13	&	(2.83$\pm$0.37)E+12	&	(1.66$\pm$0.41)E+14	&	(7.19$\pm$0.73)E+13	\\
40	&	19.9		&	(1.62$\pm$0.10)E+12	&	$>$1.81E+14		&	$>$3.66E+13	&	(6.41$\pm$0.43)E+12	&	(4.57$\pm$0.51)E+14	&	(1.31$\pm$0.06)E+14	\\
100	&	15.6		&	-			&	(3.83$\pm$0.26)E+12	&	-		&	-			&	(1.43$\pm$0.43)E+14	&	(7.43$\pm$1.43)E+12	\\
158	&	19.2		&	-			&	(4.18$\pm$0.26)E+12	&	-		&	-			&	-			&	-			\\

\hline
\end{tabular}
\end{minipage}
\end{table*}

\begin{figure}
\includegraphics[height=\linewidth0, angle=90]{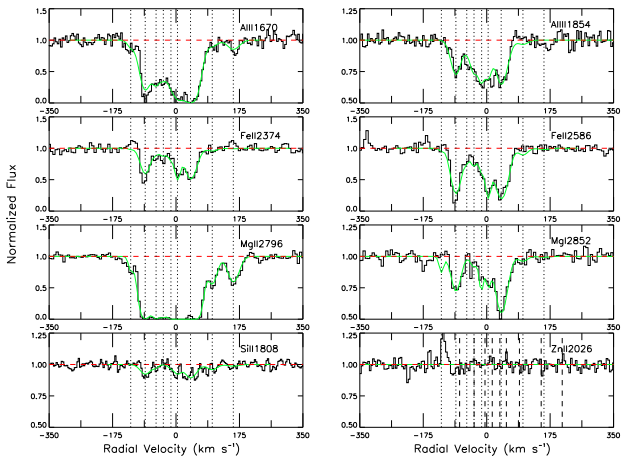}
\caption{The same as fig.1, but for the z=1.3270 system in the spectrum of Q1010+0047}
\end{figure}

\subsection{Q1224+0037 ($z_{em} = 1.482$)}

This QSO has a DLA with log N$_{\rm H \ I}$=20.88 at $z_{abs}=1.2346$ (system A) and a sub-DLA with log 
N$_{\rm H \ I}$=20.00 at $z_{abs}=1.2665$ (system B) \citep{Nest04, Rao06}. System A shows a relatively simple velocity structure with 5 components 
needed to fit the observed profile. The profile spans $\sim$129 \kms.
Strong Mg II $\lambda\lambda$ 2796, 2803 and Fe II $\lambda\lambda$ 2344, 2374, 2382 lines
were detected in this system. Zn II $\lambda\lambda$ 2026 or 2062 lines were not, however, detected in this system at S/N $\sim$ 18 in the 
region, with [Zn/H]$<$-1.62. Strong Al II $\lambda$ 1670 absorption was seen in this system along with weaker
 Al III $\lambda\lambda$ 1854, 1862 lines. 
Strong Si II $\lambda\lambda$ 1526, 1808 were also seen with log N$_{\rm Si \ II}$=15.10. The weaker Fe II $\lambda\lambda$ 2249, 2260
lines were not detected so only a lower limit could be placed on log N$_{\rm Fe \ II} > 15.11$ based on the saturated 
Fe II $\lambda\lambda$ 2344, 2374, 2382 lines. The metallicities based on Si, Fe, and Zn are [Si/H]=-1.32 [Zn/H]$<$-1.62, and [Fe/H]$>$-1.24.
The Cr II $\lambda\lambda$ 2056, 2062 lines were observed in this system with 
rest frame equivalent widths of 47 and 30 m\ang respectively. The Zn II $\lambda$ 2026 line was not however detected. 
Table 8 gives the results of the profile fitting for each of the components. Figure 6 shows velocity plots for several of the lines of 
interest in this system. 

System B shows a more complicated velocity structure with a total of 8 components needed to fit the observed profile. 
This system also shows a larger velocity width of $\sim$253 \kms.
Mg II $\lambda\lambda$ 2796, 2803 lines were detected along with strong Fe II $\lambda\lambda$ 2344, 2374, 2382 lines. This system also
has a saturated Al II $\lambda$ 1670 line with log N$_{\rm Al \ II} >$ 13.72 and unsaturated Al III $\lambda\lambda$ 1854, 1862 lines 
that were fit simultaneously to give log N$_{\rm Al \ III}$ = 12.98. The Si II $\lambda$ 1526 line appears to be blended with another feature, 
as can be seen in figure 7. We therefore give an upper limit on the column density N$_{\rm Si \ II}<$14.30 based on the non-detection of the 
Si II $\lambda$ 1808 line. There also may be some blending near the $v = -104$ \kms component in the Fe II $\lambda$ 2600 due to the lack of this component in 
the other Fe II lines. No Zn II $\lambda\lambda$ 2026, 2062 lines were seen with S/N $\sim$ 20 in the region.
Table 9 gives the results of the profile fitting for system B. Velocity plots of lines of interest are shown in figure 7.

\begin{table*}
\begin{minipage}{165mm}
\begin{center}
\caption{Same as table 3, but for the z=1.2346 DLA in Q1224+0037.}
\begin{tabular}{lccccccccc}
\hline
\hline
Vel	&b$_{eff}$	&	Mg I 			&	Mg II			&	Al II		&	Al III			&	Si II			&	Cr II			&	Fe II			\\
\hline
-60	&	8.2	&	-			&	(5.85$\pm$1.12)E+13	&	$>$2.14E+12	&	(7.01$\pm$2.02)E+11	&	-			&	-			&	(3.69$\pm$1.12)E+13	\\
-32	&	10.8	&	(5.14$\pm$0.64)E+11	&	$>$9.68E+13		&	$>$2.04E+13	&	(4.97$\pm$0.34)E+12	&	(8.45$\pm$0.94)E+14	&	(9.27$\pm$1.40)E+12	&	$>$3.69E+14		\\
-9	&	8.0	&	(8.93$\pm$0.84)E+11	&	$>$3.73E+14		&	$>$1.50E+13	&	(3.29$\pm$0.29)E+12	&	(3.19$\pm$0.75)E+14	&	(3.77$\pm$1.19)E+12	&	$>$8.51E+14		\\
16	&	10.4	&	(7.63$\pm$0.74)E+11	&	$>$1.82E+14		&	$>$2.14E+12	&	(4.44$\pm$0.32)E+12	&	(9.16$\pm$2.85)E+13	&	-			&	-			\\
37	&	8.2	&	-			&	(8.12$\pm$1.24)E+12	&	$>$1.69E+12	&	-			&	-			&	-			&	-			\\

\hline
\end{tabular}
\end{center}
\end{minipage}
\end{table*}

\begin{table*}
\begin{minipage}{115mm}
\begin{center}
\caption{Same as table 3, but for the z=1.2665 absorber in Q1224+0037.}
\begin{tabular}{lcccccccc}
\hline
\hline
	Vel	&b$_{eff}$	&	Mg I			&	Mg II		&	Al II		&	Al III			&	Fe II			\\
\hline
	-136	&	7.8	&	-			&	$>$8.95E+13	&	$>$7.11E+12	&	(1.00$\pm$0.28)E+12	&	(4.95$\pm$0.94)E+13	\\
	-104	&	10.4	&	-			&	$>$6.62E+12	&	$>$6.46E+11	&	-			&	-			\\
	-76	&	10.5	&	-			&	$>$1.13E+14	&	$>$4.21E+12	&	(2.15$\pm$0.33)E+12	&	(1.07$\pm$0.34)E+13	\\
	-26	&	5.1	&	-			&	$>$3.48E+13	&	$>$1.49E+12	&	-			&	-			\\
	0	&	8.1	&	(3.96$\pm$0.99)E+11	&	$>$1.23E+14	&	$>$7.16E+12	&	(2.75$\pm$0.36)E+12	&	(5.01$\pm$0.96)E+13	\\
	28	&	7.9	&	(5.93$\pm$1.14)E+11	&	$>$1.90E+14	&	-		&	(7.72$\pm$2.71)E+11	&	(1.45$\pm$0.30)E+14	\\
	57	&	7.9	&	-			&	$>$8.54E+13	&	$>$1.06E+13	&	(1.47$\pm$0.30)E+12	&	(6.15$\pm$1.17)E+13	\\
	92	&	8.1	&	-			&	$>$3.12E+14	&	$>$2.09E+13	&	(1.45$\pm$0.30)E+12	&	(2.82$\pm$0.56)E+13	\\

\hline
\end{tabular}
\end{center}
\end{minipage}
\end{table*}

\begin{figure}
\includegraphics[height=\linewidth, angle=90]{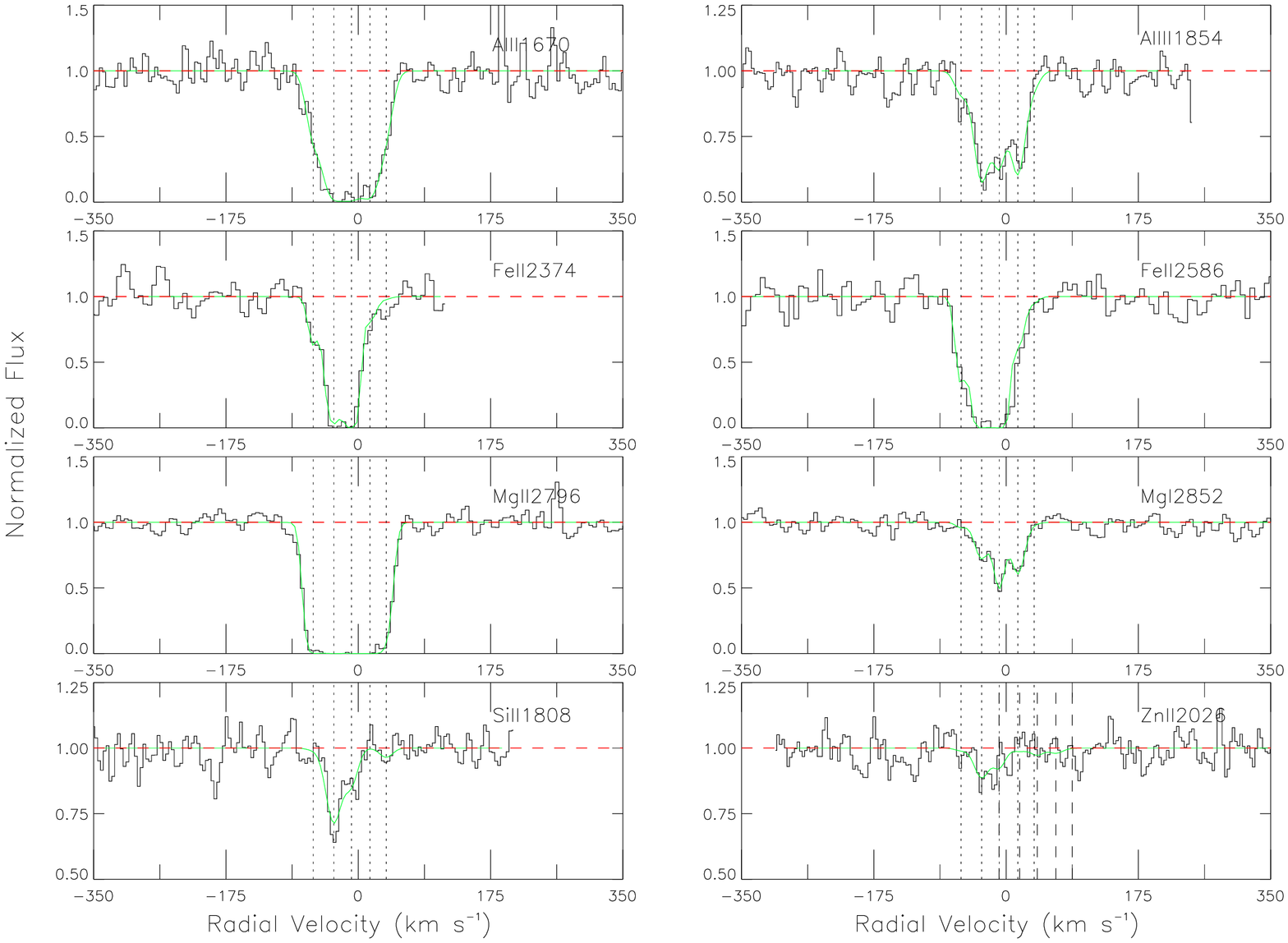}
\caption{The same as fig.1, but for the z=1.2346 system in the spectrum of Q1224+0037}
\end{figure}

\begin{figure}
\includegraphics[height=\linewidth, angle=90]{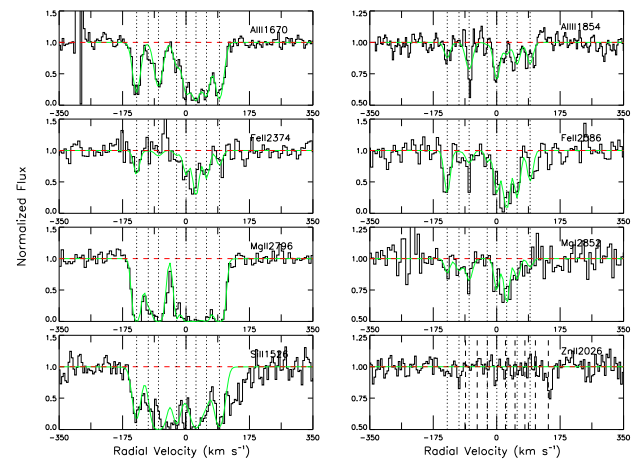}
\caption{The same as fig.1, but for the z=1.2665 system in the spectrum of Q1224+0037}
\end{figure}

\subsection{Q2331+0038 ($z_{em} = 1.714$)}
    This QSO harbors a sub-DLA with log N$_{\rm H \ I}$=20.00 \citep{Nest04, Rao06}. This system exhibits a complex velocity structure needing 9 components 
to fit the profile. The total profile spans more than 558 \kms. 
 The 2 components at $v = -255$ and $v = -204$ \kms appear mainly in the strong 
Mg II $\lambda\lambda$ 2796, 2803 and the stronger Fe II transitions. We detect Zn II $\lambda$ 2026 in this system at $\sim$
 3$\sigma$ with log N$_{\rm Zn  \ II}$ = 12.12 and [Zn/H]=-0.51. The Zn II is found solely in the $v = 45$ \kms component, which is the 
strongest component in Mg I $\lambda$ 2852 and the Fe II lines. The contribution of Mg I to the blended Zn II + Mg I $\lambda$ 2026 line
is small, W$_{0}$ $\sim$ 300  m\ang for the Mg I $\lambda$ 2852 line, well below the level discussed in $\S$ 4.1 where saturation may become important. 
There appears to be an interloper at $v \sim -170$ \kms of the Al III $\lambda$ 1854 line.
Table 10 summarizes the results of the profile fitting analysis for this system, and figure 8 shows the velocity plots for several lines of interest. 	

\begin{figure}
\includegraphics[height=\linewidth, angle=90]{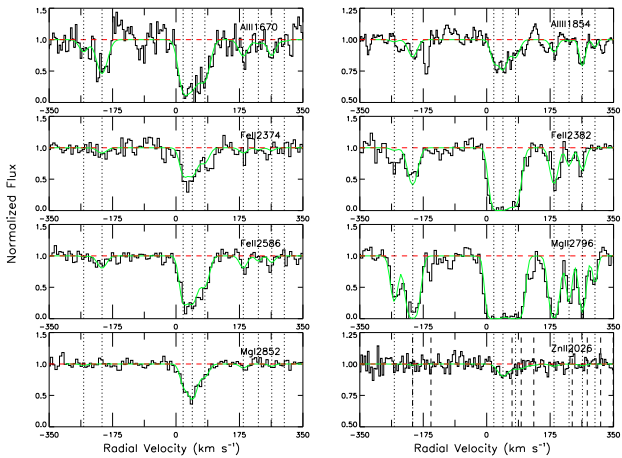}
\caption{The same as fig.1, but for the z=1.1414 system in the spectrum of Q2331+0038}
\end{figure}

\begin{table*}
\begin{minipage}{135mm}
\begin{center}
\caption{Same as table 3, but for the z=1.1414 absorber in Q2331+0038.}
\begin{tabular}{lcccccccc}
\hline
\hline
	Vel	&b$_{eff}$	&	Mg I			&	Mg II			&	Al II			&	Al III			&	Fe II		&	Zn II			\\
\hline
	-255	&	10.7	&	-			&	(7.09$\pm$0.76)E+12	&	-			&	-			&	-			&	-		\\
	-204	&	15.1	&	-			&	$>$4.43E+13		&	(3.76$\pm$0.67)E+12	&	(1.61$\pm$0.34)E+12	&	(1.42$\pm$0.22)E+13	&	-		\\
	20	&	10.4	&	(5.77$\pm$0.83)E+11	&	$>$2.62E+14		&	$>$8.24E+12		&	(1.41$\pm$0.31)E+12	&	(6.48$\pm$0.75)E+13	&	-		\\
	45	&	16.5	&	(1.93$\pm$0.13)E+12	&	$>$1.93E+13		&	$>$7.31E+12		&	(3.21$\pm$0.42)E+12	&	(1.02$\pm$0.09)E+14	&	(1.32$\pm$0.32)E+12\\
	80	&	13.8	&	(4.50$\pm$0.70)E+11	&	$>$2.21E+14		&	$>$4.74E+12		&	(1.52$\pm$0.33)E+12	&	(4.03$\pm$0.39)E+13	&	-		\\
	186	&	11.4	&	-			&	$>$3.44E+13		&	(1.26$\pm$0.40)E+12	&	(9.27$\pm$2.95)E+11	&	(1.16$\pm$0.20)E+13	&	-		\\
	227	&	7.2	&	-			&	$>$6.65E+12		&	-			&	-			&	-			&	-		\\
	264	&	8.3	&	-			&	$>$1.35E+13		&	(1.13$\pm$0.41)E+12	&	(1.74$\pm$0.31)E+12	&	(5.12$\pm$0.16)E+12	&	-		\\
	299	&	5.0	&	-			&	(2.54$\pm$0.49)E+12	&	-			&	-			&	-			&	-		\\

\hline
\end{tabular}
\end{center}
\end{minipage}
\end{table*}

\section{Results}
\subsection{Total Column Densities}
   In table 11 we summarize the total column densities for the different species for all of the absorbers in this sample. 
Cells with entries ``$\cdots$'' have undetermined column densities because of  lack of wavelength 
coverage, being blended with Lyman-$\alpha$ forest clouds, blending with atmospheric absorption 
absorption bands, or very high noise levels due to the inefficiencies  of the spectrograph at the wavelength extremes. 
Several systems showed significant saturation in the lines of Mg II, Al II, and Fe II so the values listed are given as lower 
limits. In general the column densities derived from the AOD method are slightly less than the ones derived from the profile fitting 
analysis for unsaturated lines, and differ by up to $\sim$ 0.8 dex in the saturated Mg II lines. 
For stronger absorption lines, \citet{Sav91} showed that the instrumental resolution affects the column density derived via the apparent optical depth method.
For fully resolved profiles, \citet{Sav91} showed that the apparent column density was very close to the ``true'' column density, whereas for unresolved 
profiles, even in moderately strong lines there was significant departure from the ``true'' column density that indicate hidden absorption in the profile. 
At the $\sim$20 \kms resolution obtained with the settings used on MIKE, there are likely unresolved components and therefore departures from the true
column density by those given from the apparent optical depth method. 

In table 12 we give the metallicity of these systems from profile fitting, and also the
abundance ratio [Fe/Zn], which is often taken as an indicator of dust depletion. We also give [Si/Fe], [Ca/Fe], [Cr/Fe], and [Mn/Fe] for these systems.
The ratios of column densities of the adjacent ions Mg II / Mg I and Al II / Al III are given, as well as Mg II / Al III and Fe II / Al III. For every absorber 
in this sample, there is an under-abundance of Mn relative to Fe. This could be a sign of the even-odd effect, although dust depletion can confuse this. 
There also appears to be a trend similar to that of DLA systems of higher depletion of Fe with increasing metallicity \citep{PW02,Led03,HF06,Mei06} seen in these sub-DLA 
systems. Only the $z=1.3270$ sub-DLA in Q1010-0047 shows any signs of $\alpha$-enhancement with [Si/Fe]=+0.42 although dust depletion is also a possibility. 
It is also worth noting that this system shows
a low metallicity based on Zn ([Zn/H]$<$-0.75), but higher [Fe/H] and [Si/H]. This would suggest that, at least for this system, comparing Si and Zn 
metallicities may give vastly different results. This type of non-standard abundance may not be unexpected, the zones sampled by the background 
QSOs are small due to the small size of the QSO beam, and multiple regions probed within a galaxy can show up as different components. 
Similarly, contamination by a small number of supernovae may produce odd abundance patterns due to the relatively young, unmixed gas.

\begin{table*}
{\scriptsize
\caption{Total column densities from the absorbers in this sample.}
\begin{tabular}{lcccccccccccc}
\hline
\hline	
QSO		&$z_{abs}$	&log N$_{\rm H \ I}$ 	&log N$_{\rm Mg \ I}$	&log N$_{\rm Mg  \ II}$	&log N$_{\rm Al \ II}$	&log N$_{\rm Al \ III}$	&log N$_{\rm Si \ II}$&	log N$_{\rm Ca \ II}$	&log N$_{\rm Cr \ II}$	&log N$_{\rm Mn \ II}$	&log N$_{\rm Fe \ II}$	&	log N$_{\rm Zn \ II}$	\\
		&		&	cm$^{-2}$ 	&	cm$^{-2}$ 	&	cm$^{-2}$ 	&	cm$^{-2}$ 	&	cm$^{-2}$ 	&	cm$^{-2}$ 	&	cm$^{-2}$ 	&	cm$^{-2}$ 	&	cm$^{-2}$ 	&	cm$^{-2}$ 	&	cm$^{-2}$ 	\\
\hline	 		 		 		 		 		 		 		 		 		 		 		 		 
Q0354-2724	&	1.4051	&	20.18$\pm$0.15	&	12.72$\pm$0.02	&	$>$15.08	&	$\cdots$	&	$\cdots$	&	$\cdots$	&	$\cdots$	&	13.35$\pm$0.06	&	12.90$\pm$0.06	&	15.15$\pm$0.05	&	12.73$\pm$0.03	\\
AOD		&		&			&	12.70$\pm$0.01	&	$>$14.39	&			&			&			&			&	13.25$\pm$0.04	&	12.82$\pm$0.03	&	15.03$\pm$0.01	&	12.72$\pm$0.03	\\
Q0826-2230	&	0.9110	&	19.04$\pm$0.04	&	12.09$\pm$0.03	&	$>$13.76	&	$\cdots$	&	$<$11.72	&	$<$14.22	&	11.75$\pm$0.06	&	$<$12.11	&	$<$11.37	&	13.57$\pm$0.04	&	12.35$\pm$0.07	\\
AOD		&		&			&	12.06$\pm$0.02	&	$>$13.71	&			&			&			&	11.42$\pm$0.04	&			&			&	13.43$\pm$0.12	&	12.41$\pm$0.05	\\
Q1009-0026	&	0.8426	&	20.20$\pm$0.06	&	11.73$\pm$0.09	&	$>$14.30	&	$\cdots$	&	12.73$\pm$0.05	&	$\cdots$	&	$\cdots$	&	$<$12.41	&	12.26$\pm$0.05	&	14.39$\pm$0.04	&	$<$11.85	\\
AOD		&		&			&	11.80$\pm$0.05	&	$>$13.87	&			&	12.74$\pm$0.03	&			&			&			&	12.28$\pm$0.04	&	14.37$\pm$0.03	&			\\
Q1009-0026	&	0.8866	&	19.48$\pm$0.05	&	12.41$\pm$0.04	&	$>$14.52	&	$\cdots$	&	13.12$\pm$0.03	&	$<$14.26	&	12.26$\pm$0.04	&	$<$12.11	&	$<$11.37	&	14.58$\pm$0.04	&	12.36$\pm$0.04	\\
AOD		&		&			&	12.43$\pm$0.02	&	$>$14.32	&			&	13.00$\pm$0.02	&			&	12.10$\pm$0.04	&			&			&	14.33$\pm$0.06	&	12.38$\pm$0.04	\\
Q1010-0047	&	1.3270	&	19.81$\pm$0.05	&	12.46$\pm$0.02	&	$>$15.17	&	$>$13.87	&	13.28$\pm$0.02	&	15.02$\pm$0.02	&	$\cdots$	&	$<$12.37	&	$<$11.85	&	14.53$\pm$0.03	&	$<$11.69	\\
AOD		&		&			&	12.49$\pm$0.02	&	$>$14.26	&	$>$13.72	&	13.29$\pm$0.02	&	14.86$\pm$0.15	&			&			&			&	14.50$\pm$0.02	&			\\
Q1224+0037	&	1.2346	&	20.88$\pm$0.05	&	12.34$\pm$0.04	&	$>$14.86	&	$>$13.62	&	13.13$\pm$0.04	&	15.10$\pm$0.07	&	$\cdots$	&	13.12$\pm$0.09	&	$<$12.14	&	$>$15.11	&	$<$11.89	\\
AOD		&		&			&	12.31$\pm$0.03	&	$>$14.30	&	$>$13.59	&	13.07$\pm$0.03	&	15.01$\pm$0.05	&			&	13.14$\pm$0.05	&			&	$>$14.99	&			\\
Q1224+0037	&	1.2665	&	20.00$\pm$0.07	&	12.00$\pm$0.10	&	$>$14.97	&	$>$13.72	&	12.98$\pm$0.08	&	$<$14.30	&	$\cdots$	&	$<$12.37	&	$<$12.16	&	14.54$\pm$0.09	&	$<$11.85	\\
AOD		&		&			&	12.21$\pm$0.08	&	$>$14.25	&	$>$13.56	&	12.74$\pm$0.14	&			&			&			&			&	14.36$\pm$0.04	&			\\
Q2331+0038	&	1.1414	&	20.00$\pm$0.05	&	12.48$\pm$0.05	&	$>$14.79	&	$>$13.42	&	13.02$\pm$0.09	&	$<$14.33	&	$\cdots$	&	$<$12.37	&	$<$11.79	&	14.38$\pm$0.03	&	12.12$\pm$0.11	\\
AOD		&		&			&	12.47$\pm$0.03	&	$>$14.39	&	$>$13.33	&	12.86$\pm$0.14	&			&			&			&			&	14.44$\pm$0.05	&	12.22$\pm$0.09	\\

\hline
\end{tabular}
}
\end{table*}

\begin{table*}
{\scriptsize
\begin{minipage}{180mm}
\caption{Abundances for the absorbers in this sample.}
\begin{tabular}{lcccccccccccc}
\hline
\hline	
QSO		&	$z_{abs}$&	[Zn/H]		&	[Fe/H]		&	[Fe/Zn]		&	[Si/Fe]		&	[Ca/Fe]		&		[Cr/Fe]	&	[Mn/Fe]		&	Al III / Al II	&Mg II / Mg I	&	Mg II / Al III	&	Fe II / Al III	\\
\hline
$[X/Y]_{\sun}$	&		&	-7.37		&	-4.53		&	+2.85		&	+0.07		&	-1.13		&	-1.82		&	-1.97		&			&		&			&			\\
\hline																							
Q0354-2724	&	1.4051	&	-0.08$\pm$0.16	&	-0.50$\pm$0.16	&	-0.43$\pm$0.06	&	$\cdots$	&	$\cdots$	&	0.02$\pm$0.07	&	-0.35$\pm$0.08	&	$\cdots$	&	$>$2.36	&	$\cdots$	&	$\cdots$	\\
Q0826-2230	&	0.9110	&	+0.68$\pm$0.08	&	-0.94$\pm$0.06	&	-1.63$\pm$0.08	&	$<$0.58		&	-0.69$\pm$0.07	&	$<$+0.36	&	$<$-0.23	&	$\cdots$	&	$>$1.67	&	$>$2.04		&	$>$1.85		\\
Q1009-0026	&	0.8426	&	$<$-0.98	&	-1.28$\pm$0.07	&	$>$-0.31	&	$\cdots$	&	$\cdots$	&	$<$-0.16	&	-0.16$\pm$0.06	&	$\cdots$	&	$>$2.57	&	$>$1.57		&	1.66$\pm$0.06	\\
$\cdots$	&	0.8866	&	+0.25$\pm$0.06	&	-0.37$\pm$0.06	&	-0.63$\pm$0.06	&	$<$-0.39	&	-1.19$\pm$0.06	&	$<$-0.65	&	$<$-1.24	&	$\cdots$	&	$>$2.11	&	$>$1.40		&	1.46$\pm$0.05	\\
Q1010-0047	&	1.3270	&	$<$-0.75	&	-0.75$\pm$0.06	&	$>$0.01		&	0.42$\pm$0.07	&	$\cdots$	&	$<$-0.34	&	$<$-0.71	&	$<$-0.71	&	$>$2.71	&	$>$1.69		&	1.25$\pm$0.05	\\
Q1224+0037	&	1.2346	&	$<$-1.62	&	$>$-1.24	&	$>$0.37		&	$<$-0.08	&	$\cdots$	&	$<$-0.17	&	$\cdots$	&	$<$-0.49	&	$>$2.52	&	$>$1.73		&	$>$1.98		\\
$\cdots$	&	1.2665	&	$<$-0.78	&	-0.93$\pm$0.11	&	$>$-0.16	&	$<$-0.31	&	$\cdots$	&	$<$-0.35	&	$<$-0.41	&	$<$-0.74	&	$>$2.97	&	$>$1.99		&	1.56$\pm$0.12	\\
Q2331+0038	&	1.1414	&	-0.51$\pm$0.12	&	-1.09$\pm$0.06	&	-0.59$\pm$0.11	&	$<$-0.12	&	$\cdots$	&	$<$-0.19	&	$<$-0.62	&	$<$-0.40	&	$>$2.31	&	$>$1.77		&	1.36$\pm$0.09	\\

\hline
\end{tabular}
\end{minipage}
}
\end{table*}

\subsection{The Al III to Al II Ratio and Ionized Gas}

    Based on the ionization potentials of the Al$^{0}$ and Al$^{+}$ ions (5.99 and 18.83 eV respectively), one would expect that the majority of the
Al seen in DLA and sub-DLA systems would be in the first ionization stage (Al$^{+}$), and that the second ionization stage (Al$^{++}$)
would be all but absent due to the shielding of photons with energies greater than the ionization potential of hydrogen (13.6 eV). 
The Al III $\lambda\lambda$ 1854, 1862 
lines are nonetheless often seen in DLA spectra although the majority is always in Al II.
 As is seen in most DLAs, the Al III $\lambda\lambda$ 1854, 1862 lines of the systems studied here 
have very similar component structures to the lower ionization state lines such as Fe II, Mg II. Higher ionization state transitions such as C IV or S IV 
often show very different component structures than the low ion transitions, implying pockets of local, highly ionized gas within
the DLA/sub-DLA, or possibly galactic halos. The similar structure of the Al III and low ion profiles suggests that these systems are physically connected.
	
	The Al III / Al II ratios have in the past been used to investigate the ionization states of QSO absorbers. 
\citet{Vlad01} observed an anti-correlation between the Al III / Al II ratio and N$_{\rm H \ I}$ based on a sample of 20 DLAs, where the Al II
column density was estimated from Si II in most cases. They suggested that the higher N$_{\rm H \ I}$ systems also had lesser degrees of ionization. 
It should be noted however that there may be issues with using Si as a surrogate for Al \citep{Bark84}.
\citet{Des03} noted however that when the data were extended into the column density region of sub-DLAs, the trend seemed to disappear. 
In figure 9, we show the Al III / Al II ratio for the absorbers from this work, \citet{Des03}, and \citet{Vlad01}.
As can be seen from the original data of \citet{Vlad01} (blue squares, black diamonds, and purple triangle), the before mentioned trend 
does seem to hold for these data. It can also be seen in figure 9 that the $z > 1.8$ sub-DLAs of \citet{Des03} and the $z < 1.5$ absorbers seen
here show similar Al III / Al II ratios, suggesting little evolution of this parameter although the sample sizes are still small. 
With one exception of 30 cases in figure 9, which covers 19$<$log N$_{\rm  H \ I}$$<$21.7, N$_{\rm Al \ III}$/N$_{\rm Al \ II}\la$-0.25. 
Models from \citet{Petit94} also showed little change in the Al III/Al II ratio especially when 17 $\la log N_{\rm H \ I} \la$ 20.
Similarly, \citet{York06} found the ratio to change very little over a wide range of reddenings corresponding to the same range in N$_{\rm  H \ I}$.

\begin{figure}
\vspace{8mm}
\includegraphics[width=\linewidth]{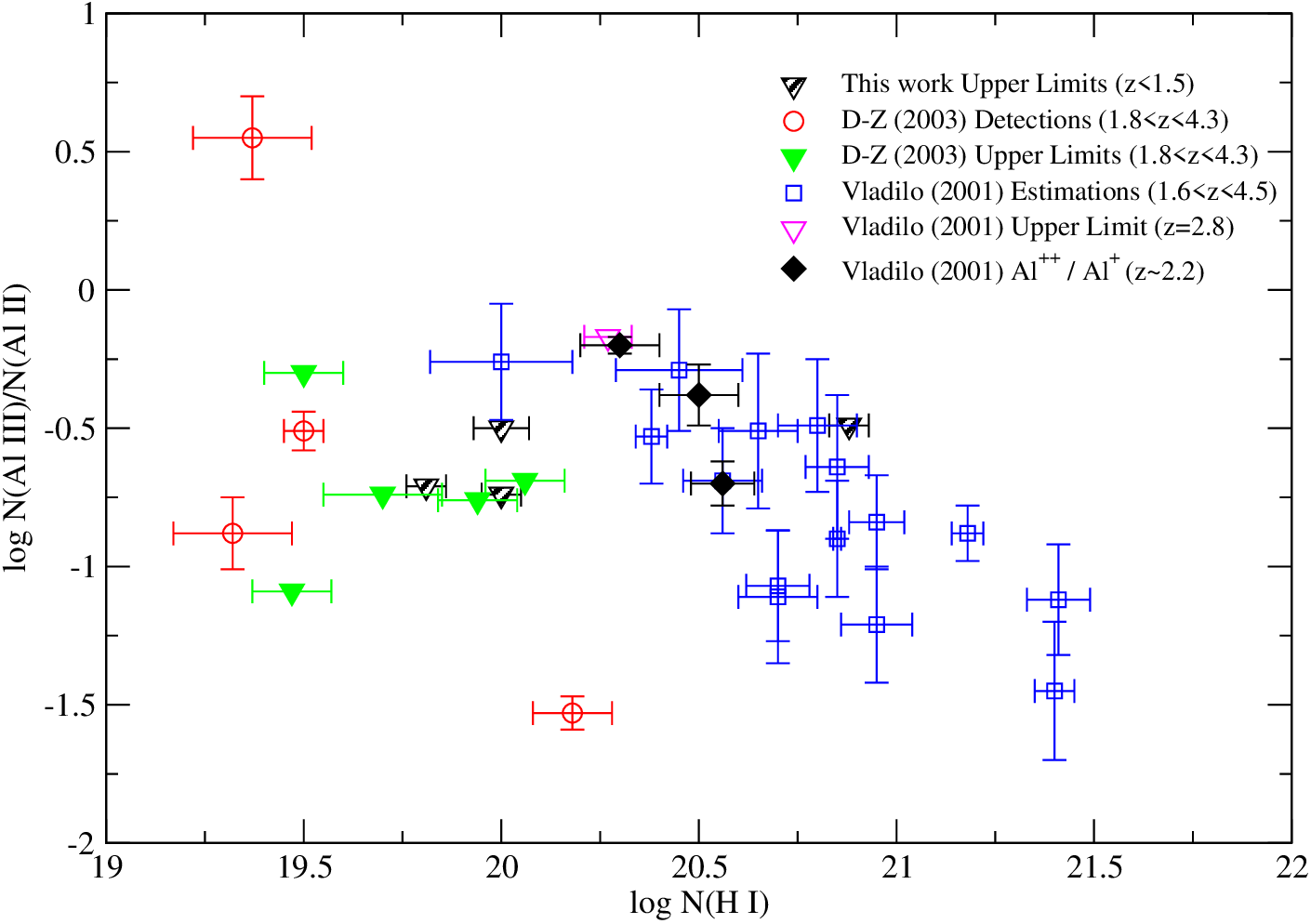}
\vspace{3mm}
\caption{The Al III / Al II ratio for the absorbers in this sample (striped triangles), \citet{Des03} 
(solid triangles), and \citet{Vlad01} (blue squares, purple triangle, and black diamonds). 
For the \citet{Vlad01} systems, the blue squares represent systems where Al II was estimated from Si II, the black diamond points are for systems with
true Al III / Al II values, and the purple triangle is an upper limit based on Al III / Al II. 
All of the absorbers from this work with Al II measurements showed significant saturation so only lower limits could be placed on their column densities. }
\end{figure}

\subsection{Cloudy Modeling and Ionization Corrections}
It is generally assumed for DLAs that the gas has a very low ionization fraction, and that most metal ions are mainly in the first ionization state. 
This is because of the self-shielding of photons with h$\nu$ $>$ 13.6 eV due to the large cross section for ionization of H at these energies. Therefore, the
abundances reported for QSO absorbers are typically derived from lines originating from ions and may not reflect the true abundances.      
The ionization correction factor, defined here as
	$$ \epsilon=[X/H]_{\rm total} - [X^{+}/H^{0}] $$ 
where the total column densities include contributions from all ionization stages, has been investigated for DLAs by several groups 
\citep{Howk99, Vlad01, Pro02}. Their conclusions were all roughly the same; that the ionization corrections for most elements
were $\la$0.2 dex in most cases. Due to the smaller amount of HI in the sub-DLA systems, one might expect that they would show a greater amount of 
ionization. 

	\citet{Des03} showed the ionization corrections for their sample of sub-DLA at $z > 1.8$ were, in general 
small, with $<$0.2 dex corrections based on predictions of the Cloudy software package. The only absorber in their sample that 
showed significant signs of ionization effects was the log N$_{\rm H \ I}$=19.37 sub-DLA in the spectrum of Q2155+1358 at 
$z_{abs} = 3.565.$ The absorbers in the spectra of Q0826-2230 with log N$_{\rm H  \ I}=19.04$, Q1009-0026 with log N$_{\rm H \ I}$=19.48,
and Q1010-0047 with log N$_{\rm H \ I}=19.81$
have the lowest column densities in our sample, and thus might be expected to require some ionization corrections.

	To investigate further, we used the Cloudy software package to compute photo-ionization models assuming local thermodynamic equilibrium (LTE). 
Calculations were performed with version C06.02.b of Cloudy, last described by \citet{Fer98}. Grids of Cloudy models were computed assuming 
that the spectrum of ionizing radiation striking the cloud followed the form of the extra-galactic UV background of Haardt $\&$ Madau \citep{HM96, MHR99} 
at the appropriate redshift of the absorber, plus the model stellar atmosphere of Kurucz with a temperature of 30,000K to simulate a 
radiation field produced via an O/B-type star. Plots of both types of spectra can be found in Hazy, the documentation for Cloudy. 
The extra-galactic spectrum of Haardt $\&$ Madau is a harder spectrum than the Kurucz spectrum, which produces few photons with h$\nu >$2 Ryd. 
In fact, the Haardt $\&$ Madau spectrum produces a total of $\sim$ 21.13 times as many photons with h$\nu>$13.6 eV than the Kurucz spectrum. From 1-2 Ryd
though, the total ionizing photons is almost identical for both spectra. 
It has recently been shown that the contribution from local sources to the ionization of DLA systems 
is likely non-negligible compared to the ionizing background radiation field  \citep{Sch06}.  We include the radiation field of the Kurucz stellar 
spectrum in these Cloudy simulations, but do not allow that intensity to vary relative to QSOs. 
In addition, our Cloudy models have included both cosmic ray and cosmic microwave backgrounds. These simulations
however do not include radiation from local shocks caused by supernovae, or compact sources such as white dwarfs or compact binary systems, all of which 
may contribute significantly to the ionizing radiation field. Explorations into contributions of various galactic sources is beyond the scope of this work. 
	
	For each of the grids of models, the ionization parameter defined by
	$$ U=\frac{n_{\gamma}}{n_{H}}=\frac{\Phi_{912}}{cn_{H}}, $$
(where $\Phi_{912}$ is the flux of radiation with h$\nu$ $>$ 13.6 eV) was increased from log U=-6.0 to 0. Each model was tailored 
to match the N$_{\rm H \ I}$, metallicity based on Zn II, and redshift of the observed system. Due to the saturation of the strong 
Al II $\lambda$ 1670 transition, for each of the three systems where we have detections 
of Al III $\lambda\lambda$ 1854, 1862 lines only an upper limit can be placed on the ionization parameter U. Also, we are not at all able to 
constrain the shape of the ionizing spectrum that is illuminating the cloud. The possibilities presented here are but a myriad of other 
possible shapes of the spectrum of ionizing radiation. Nonetheless, using standard assumptions, outlined above, some conclusions can be drawn.

Cloudy models were calculated for all of the systems observed. We present the results of the most interesting systems here.
With log N$_{\rm H \ I}$ = 19.04 the weak sub-DLA toward Q0826-2230 may be expected to require ionization corrections. 
The only element observed in this system with
multiple ionization stages was Mg. Due to the saturation of the Mg II $\lambda\lambda$ 2796, 2803 lines only a lower limit could be placed on
the Mg$^{+}$ to Mg$^{0}$ ratio, and thus a lower limit on the ionization parameter U. 
However, the Mg II $\lambda$ 2803 line was not optically thick even in its strongest components, so
the column density given in table 12 may not be far from true. To see if the true column density could be significantly higher, we increased the column
densities in the components by a factor of 10, at which point there was significant deviation between the observed and theoretical profiles. Also, the 
column densities derived from the AOD method and profile fitting agreed to within 0.05 dex. 
Nonetheless, as can be seen from figure 10, the ionization corrections to
Zn$^{+}$ are $\la$0.2 dex even for log U$\la$-3.0 which is the typical upper limit for other sub-DLA systems observed. 
We have therefore not included any ionization correction factor into the quoted Zn abundance in table 13.

In figure 10, the horizontal lines in the center panels indicates the observed limits or ratios of the Mg II to Mg I, Al III to Al II, 
and Al III to Fe II column densities. The vertical lines are therefore the corresponding ionization parameters based on these values. Based on the Mg II to Mg I
ratio in the $z_{abs}=0.9110$ sub-DLA in the spectrum of Q0826-2230, the ionization parameter is a lower limit due to the saturation of the Mg II lines. 

The log N$_{\rm H \ I}$ = 19.48 system in Q1009-0026 may also require ionization corrections. The Al II $\lambda$ 1670 line was below the wavelengths
accessible with the MIKE spectrograph, but we the Al III $\lambda\lambda$ 1854, 1862 lines were observed. We were therefore able to constrain the ionization
parameter based on the Al III to Fe II ratio, which was log(Al III / Fe  II)=-1.66, and consequently based upon the Cloudy models log U=-3.70. The
ionization correction for Zn$^{+}$ based on this ionization parameter is small ($\sim$ 0.15 dex), so no correction corrections were introduced into table 13.

The log N$_{\rm H \ I}$ = 19.81 system in the spectrum of Q1010-0047 is another system with relatively low N$_{\rm H \ I}$ that may require ionization 
corrections. For this system, both the Al II $\lambda$ 1670 and the Al III $\lambda\lambda$ 1854, 1862 lines were observed. The Al II $\lambda$ 1670 line
is saturated, so only an upper limit could be placed on the Al${++}$ to Al$^{+}$ ratio. From the Cloudy models, the ionization parameter log U$\la$-4.15. 
Again, the ionization corrections for Zn$^{+}$ were small for values within the range of possible values for U, so no ionization correction
was added to the values in table 13.

In general, from results of the Cloudy modeling the ionization parameter U is 
typically small, log U$\la$-3.0 for the systems that had detections of the Al II $\lambda$ 1670 and Al III $\lambda\lambda$  1854, 1862 lines. 
We note, again, that the ionization parameter does depend strongly on the shape of the ionizing spectrum. 

It has been mentioned before that the Al III to Al II ratio may not be an accurate predictor of the ionization state of the gas observed in QSO 
absorption line systems \citep{Pro02}. One of the reasons for this is the uncertainties in the atomic data for Al. It has been suggested that the 
Al III to Al II ratio may provide an estimate of the ionization of the gas in DLAs to first order, but better 
diagnostic indicators of the ionization state of the gas are the Fe$^{++}$ to Fe$^{+}$ and N$^{+}$ to N$^{0}$ ratios. 
For the absorbers presented here the lines of these ions are well in the UV therefore could not be observed with our ground based observations. Therefore the 
Al III to Al II ratio is the only opportunity available to study the ionization state of the gas in these sub-DLA systems. Even if we cannot derive the 
ionization parameter absolutely, it appears that the ionization of the components does not change over a wide range of integrated column densities and 
redshifts. It should be noted that for the systems in which Al II was not observed there is no indication that Al$^{++}$ is the dominant ion. As can
be seen in table 14, the Mg II to Al III ratio is similar for systems with and without Al II measurements. If Al III were relatively higher in one system
than the others, the limit of Mg II / Al III would be lower than the others. 
r
The use of the Al III / Fe II ratio in determining the ionization parameter may introduce some uncertainties. While Cloudy assumes solar abundances, 
differential depletion between Al and Fe
may lead to an underestimation of the Al III / Fe ratio. In the Galaxy, Al is sometimes seen to be more depleted than Fe \citep{Bark84}. 
Another source of uncertainty are nucleosynthetic mismatches between these two elements. The use of column densities from adjacent ions such as Al III and Al II, 
or Mg II / Mg I assumes that the ions coexist in the gas from which these lines arise. This is almost universally true in the case of Al III and Al II, 
the line profiles are very similar. The line profiles for Mg II and Mg I are not always as similar, especially in the components at higher radial velocities 
where Mg II lines are often seen without any components seen in the Mg I profile. The ``core" components however typically do match very well, as can be seen in 
figures 1-8.

\begin{figure}
\includegraphics[width=\linewidth]{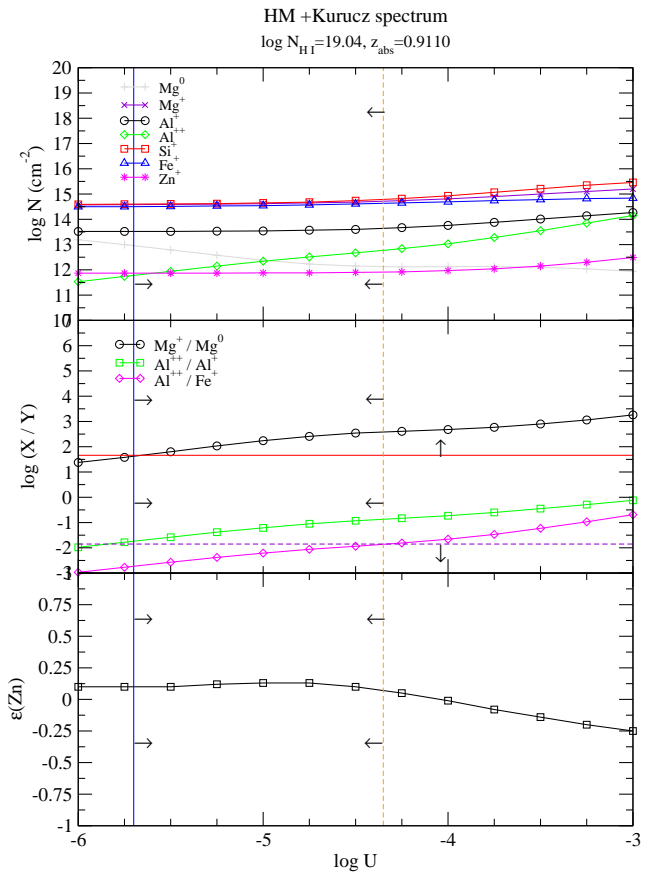}
\caption{Results of the Cloudy simulations for the $z=0.9110$ sub-DLA with log N$_{\rm H \ I}$=19.04 in the spectrum of Q0826-2230.
	 The upper panel shows the column densities for several species of interest. The middle panel shows the logarithmic ratio of 
	 Mg$^{+}$ to Mg$^{0}$, Al$^{++}$/Al$^{+}$, and Al$^{++}$/Fe$^{+}$. Here, the horizontal solid red line indicates 
	the lower limit to the observed ratio of the column densities derived from the Mg I $\lambda$ 2852 line, and the Mg II 
	$\lambda\lambda$ 2796, 2803 doublet. The vertical solid blue line is then the lower limit of the ionization parameter U, based upon the Cloudy modeling.
	The dashed lines show the observed Al III to Fe II ratio (purple), and the corresponding upper limit to the ionization parameter (yellow). 
	The lower panel shows the ionization correction factor for Zn$^{+}$ in dex.}
\end{figure}

\subsection{Sub-DLAs and the velocity dispersion-metallicity relationship}
	Based on a sample of $\sim$ 53,000 star-forming galaxies at z$\sim$0.1 observed in imaging and spectroscopy in the SDSS, 
\citet{Trem04} discovered a mass-metallicity relationship for these galaxies. Specifically, they found a correlation between stellar
mass and metallicity that spans over 3 orders of magnitude in stellar mass and one order of magnitude in metallicity. 
\citet{Nest03} noticed a relation between the width of Mg II and the metallicity, and suggested that the Mg II line width might be an indicator 
that the gas was in a deep potential well.
Evidence has recently been provided for the possible existence of a mass metallicity relationship for DLA absorbers, 
assuming the velocity width of optically thin lines to be proportional to the mass \citep{Led06}. 
As a proxy for the stellar mass of these systems, which has been difficult to detect, the velocity width is used 
as an indicator of galaxy mass, as it potentially probes the depth of the gravitational potential well of the DLA systems. 
\citet{Bou06} however find an anti-correlation between the Mg II equivalent width, and the estimated halo mass based upon an indirect mass indicator. 

	Following the analysis of \citet{Led06}, we performed an analysis of the apparent optical depth (see e.g., \citet{Sav96}) of these sub-DLA 
systems. The apparent optical depth is defined as
$$\tau(\lambda)_{a}=ln[I_{0}(\lambda)/I_{obs}(\lambda)],$$
where I$_{0}(\lambda)$ is the continuum level, and I$_{obs}(\lambda)$ is the observed intensity.
Specifically, we defined the beginning of the absorption line systems profile as the point where the apparent optical depth of the line
reached a value of 3$\sigma$ higher than the average noise value in the continuum. Similarly, the upper edge was defined as the point where 
the apparent optical depth dropped below the 3$\sigma$ level of the continuum. In figures 11 and 12, we give the apparent optical depth of the 
Fe II $\lambda$ 2374 lines vs. the radial velocity widths defined here for the systems in this sample. 
	
In figure 13 we plot the metallicity vs velocity for
the DLA absorbers from \citet{Led06} (open circles); sub-DLAs from \citet{Led06} (open squares); sub-DLAs from 
\citet{Per06a}, \citet{Per06b}, $\&$ \citet{Pro06} (filled squares);
DLAs from \citet{Per06a} $\&$ \citet{Per06b} (filled circles); the Zn detections from this work (filled stars); 
and the upper limits from this work (downward arrows).
In their analysis, \citet{Led06} used weaker features, where at most 40$\%$ of the continuum level was absorbed by the line. Due to the 
unavailability of the spectra for the sub-DLAs in the literature, we were unable to measure the velocity width directly, and the value plotted is the velocity
width given in the reference. In the case of the two sub-DLAs in \citet{Pro06}, the velocity width was estimated from the plots of the Si II $\lambda$ 1808 and
Fe II $\lambda$ 1608 lines. For the systems in our sample and \citet{Per06a,Per06b}, in addition to the Fe II $\lambda$ 2374 line we also measured 
the velocity width of the Mg II $\lambda$ 2796 or $\lambda$ 2803 line, 
which is also plotted for each system in figure 15, with a line connecting the two values. 
We have chosen these two lines for this investigation for the several reasons. Firstly, with weaker features, the velocity 
width becomes sensitive to the S/N. Secondly, the Fe II $\lambda$ 
2374 line looks much like Ca II in the SMC, so ground based work with Ca II in lower $z$ systems can be compared directly with the Fe II $\lambda$ 2374 line
in higher $z$ systems. Lastly, the stronger Mg II $\lambda$ 2796, 2803 lines represent a maximum velocity width to the line, and also include the components at
higher radial velocities that are often not seen in the intrinsically weaker features.

As can be seen from figure 15, the systems from this sample are in agreement with the general trend of the data. 
More interesting though is the fact that the sub-DLAs from this sample and
from \citet{Pro06, Per06a} are all at the highest end of the metallicity range. Indeed, the only points that lie above solar metallicity are 
those from sub-DLAs. Is this possible evidence that sub-DLAs are more representative of massive spiral or elliptical galaxies than their DLA 
counterparts? See also \citet{Kh06} for a discussion of the nature of QSO absorbers.

Because the absorbers chosen in our sample were chosen partly because of their strong Mg II or Fe II lines, it is possible that these systems would
be disposed to lie in the upper right corner of figure 13, with high metallicities and larger velocity widths. It was, however, shown that W$_{\rm Mg II}$ 
does not correlate with [Zn/H] \citep{Kul07} based on individual measurements. The isolation of points in this region is therefore a real effect.

\begin{figure}
\includegraphics[height=\linewidth, angle=90]{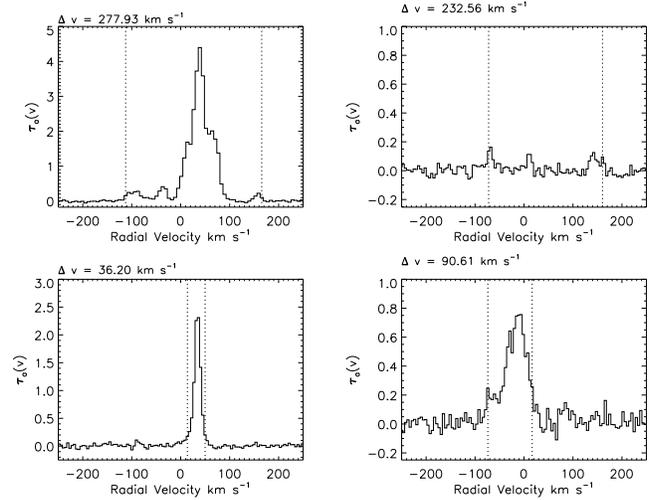}
\caption{Apparent Optical Depth plots for the Fe II $\lambda$ 2374 line for (clockwise from top left); the $z_{abs}=1.4051$ system in Q0354-2724, 
	the $z_{abs}=0.9110$ system in Q0826-2230, the $z_{abs}=0.8866$ system in Q1009-0026, and the $z_{abs}=0.8426$ system in Q1009-0026.
	The total velocity width for the system, defined in $\S$ 5.3, is given above the plot.}
\end{figure}

\begin{figure}
\includegraphics[height=\linewidth, angle=90]{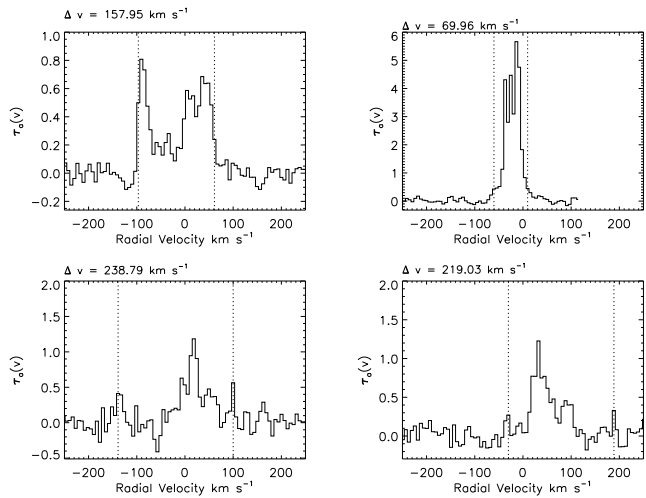}
\caption{Same as figure 13, but for the $z_{abs}=1.3270$ system in Q1010-0047, the $z_{abs}$=1.2346 system in Q1224+0037, the $z_{abs}$=1.1414 system
	 in Q2331+0038, and the $z_{abs}$=1.2665 system in Q1224+0037}
\end{figure}

\begin{figure}
\includegraphics[height=\linewidth, angle=90]{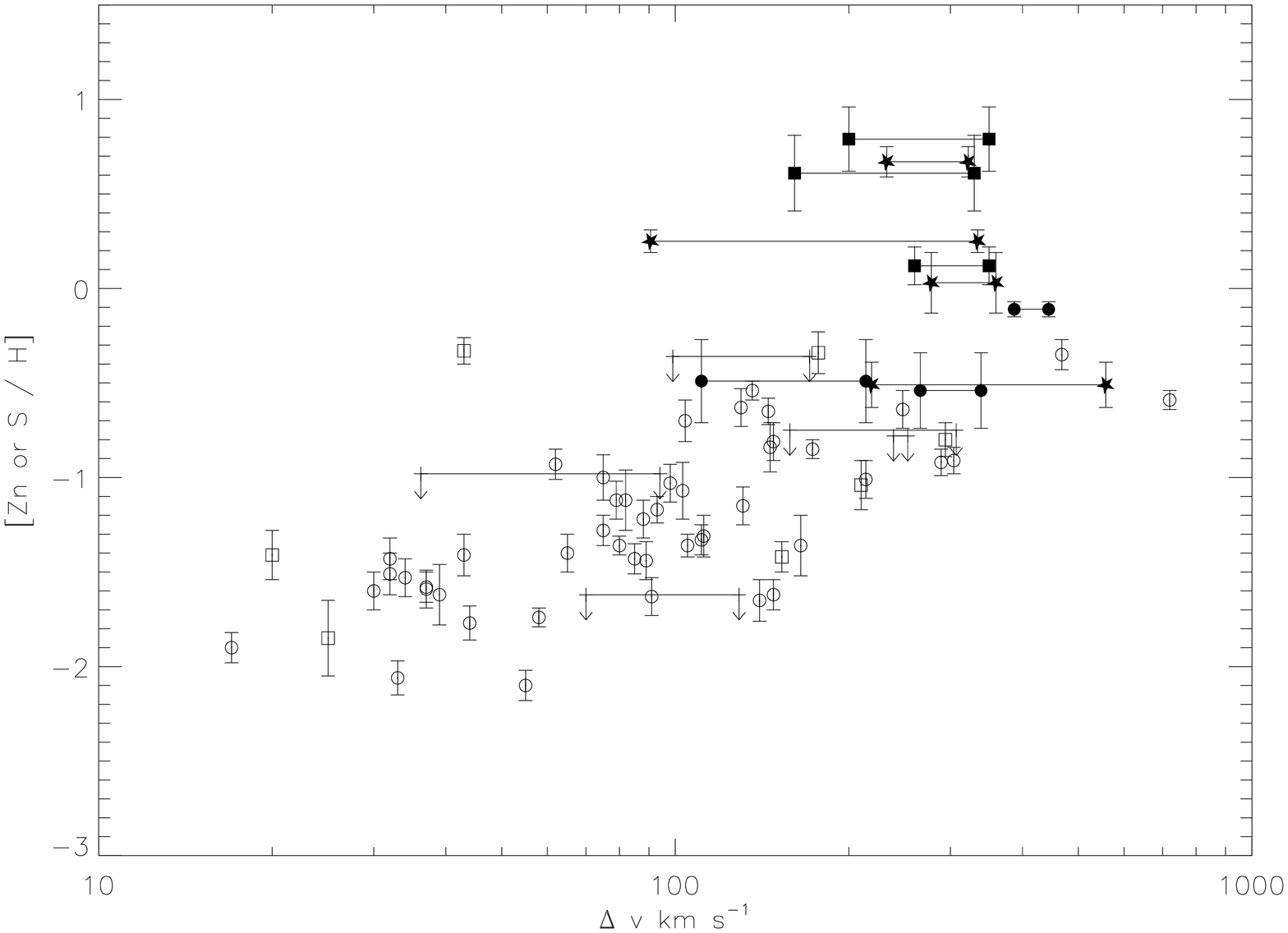}
\caption{The metallicity vs velocity width for the DLA absorbers
	from \citet{Led06} (open circles); sub-DLAs from \citet{Led06} (open squares); sub-DLAs from \citet{Per06a}, \citet{Per06b}, $\&$ \citet{Pro06}
	(filled squares); DLAs from \citet{Per06a} $\&$ \citet{Per06b} (filled circles); the Zn detections from this work (filled stars); 
	and the upper limits from this work (downward arrows).
	For the absorbers in this sample and for the absorbers in \citet{Per06a} $\&$ \citet{Per06b}, we measured the velocity width from both 
	the Fe II $\lambda$ 2374 line as seen in figures 13 and 14, and also the stronger Mg II $\lambda$$\lambda$ 2796,2803 lines. These two values 
	are plotted in figure 15 with a line connecting the two points.}
\end{figure}

\section{Conclusions and Future Work}
	In this paper, we have presented high-resolution abundance measurements of 7 sub-DLAs and 1 DLA. Several of these systems show
moderately high levels of metal enrichment compared to DLAs, with three having near solar or super-solar abundances. 
To date, the higher HI column density DLA systems with log N$_{\rm H \ I} > 20.3$ have been studied preferentially due to their high gas content 
and therefore possible high metal content. However, these systems 
show low metallicities at any redshift with [Zn/H]$\sim$-1.0 dex. Of all the (mainly neutral) QSO absorbers studied to date, all eight that have been 
observed with super-solar metallicities are all sub-DLAs \citep{Pet00, Kh04, Per06a, Pro06}. From the results of the Cloudy modeling presented here,
the ionization corrections for these absorbers in general appear to be low, consistent with the ionization corrections found by \citet{Des03} for the sub-DLAs at
higher redshift. We note however that the shape of the ionizing spectrum did strongly effect the ionization corrections 
from the Cloudy models, so the true ionization corrections could possibly be much larger. We have based our calculations on the radiation fields currently 
thought to be most relevant. Conditions will vary from component to component, and from system to system based on their respective temperatures and densities, 
and on the proximity or not of X-ray sources, UV stellar sources, and UV emission from shocks. The input spectrum is therefore certainly a source
of uncertainty, but other input parameters ranging from atomic data to the morphology assumed by Cloudy in modeling DLA systems are 
likely sources of uncertainties.

 On the matter of the high (i.e. super-solar) abundances, the errors for all our values are based on formal errors. 
We cannot quantitatively evaluate possible sources of error from external sources outside the range of the assumptions used in the analysis  
of N$_{\rm H \ I}$ and the ionization corrections. The matter of N$_{\rm H \ I}$ determinations is discussed in the appendix, and ionization corrections are 
discussed in $\S$ 5.3. The former requires higher resolution, higher S/N UV spectra to improve the confidence in the errors, 
and the latter requires a much more detailed knowledge of the radiation field in the absorbers as well as the geometry. N$_{\rm H \ I}$ determinations could
potentially be improved when HST is refurbished with UV spectroscopic capability. The ionization corrections are more difficult to determine, and will need to
be continually improved when more is learned about the environment of the absorbers. For the moment, we urge caution when interpreting the highest
abundances, until more cases are verified or these two matters are more fully dealt with. 

 The errors reported in table 12 are based upon the errors given by \citet{Rao06}, with no ionization corrections. 
 Taking the errors for N$_{\rm H \ I}$, conservatively estimated 
in the appendix, our estimates on the ionization correction from the cloudy modelling, and the error on N$_{\rm Zn \ II}$, the metallicities for the 
$z_{\rm abs}$=0.9110 and $z_{\rm abs}$=0.8866 sub-DLAs in Q0826-2230 and Q1009-0026 are +0.68$\pm$0.17 and +0.25$\pm$0.16 respectively with the errors added in
quadrature. We have however presented arguments that the observed trends in ion ratios over large ranges of N$_{\rm H \ I}$  make our conclusions valid, for the purposes 
of this paper.

	Although progress has been made recently in the understanding of sub-DLA absorbers, more observations are needed to fully understand 
the role of sub-DLAs in the metal enrichment history of the universe. Aside from the two absorbers in \citet{Per06a, Per06b}, our measurements 
constitute the only available high resolution data available for sub-DLAs at z $<$ 1.5.  As was pointed out in \citet{Pro06}, 
these very metal rich sub-DLAs may be rare, but even if they are $\la$1$\%$ of the total population they will contribute significantly to the mean metallicity
because of the much larger number density of sub-DLA systems. Clearly, more observations are needed to determine how common metal-rich
sub-DLAs are, to see if sub-DLAs do harbor a large fraction of the missing metals, and to extend the mass-metallicity relationship for 
QSO absorbers to include more sub-DLAs. Future HST observations would allow us to better determine the ionization corrections using 
the Fe III $\lambda$ 1122, N II $\lambda$ 1084, and  N I $\lambda\lambda$ 1134, 1199 lines respectively. 
The two transitions of S III ($\lambda\lambda$ 1012, 1190) provide a more likely unblended source of lines that can be used to check for ionization 
corrections, though S III probes somewhat harder radiation fields than Fe III and N II. 

\section*{Acknowledgments}
Thanks to the anonymous referee for the helpful comments that improved this work. 
We thank the helpful staff of Las Campanas Observatory for their assistance during the observing runs. Thanks to
J.X. Prochaska for his help with the MIKE reduction software. We thank Gary Ferland for both developing the Cloudy code and 
for answering questions concerning Cloudy simulations. Thanks to S.M. Rao and D. Turnshek for discussions about fitting the Ly-$\alpha$ profiles.  
We thank Don Lamb and Jim Truran for enlightening discussions on SNe enrichment. 
J. Meiring and V.P. Kulkarni gratefully acknowledge support from the National Science 
Foundation grants AST-0206197 and AST-0607739, and the NASA/STScI grant GO-9441. 
J. Meiring acknowledges partial support from South Carolina Space Grant graduate student fellowship.

\section*{Appendix - N$_{\rm H \ I}$ determinations}
   The systems studied in this work all have known N$_{\rm H \ I}$ from HST spectra. Here we provide plots of the Voigt profiles of the
Lyman-$\alpha$ transition using the best fit values of the column density from \citet{Rao06}. Due to the low resolution and S/N of
these UV spectra, only one component was used in the fits. Several factors such as blending with Lyman-$\alpha$ forest lines, 
continuum fitting and component placement add difficulties to the profile fitting with these spectra. 
These effects are largely unknown in each case and formal errors that account for a wide range of possibilities cannot be made until there are higher 
S/N UV spectra from the refurbished Hubble Space Telescope. We show in figures 14 and 15 the Voigt profiles corresponding 
to the column densities given by \citet{Rao06} and convolved with the instrumental spread function, superimposed on the
archival data from HST cycle 6 program 6577, cycle 9 program 8569, and cycle 11 program 9382. 
We note that the normalization, i.e., the continuum fit that we define may differ from that adopted by \citet{Rao06}. For our continuum fits, a polynomial 
typically of order $\la$6 was used, and the absorption line itself was excluded from the fitting region.
Also over-plotted are profiles with H I column densities 
smaller and larger by 0.15 dex. For the sub-DLAs in our sample, the plots are not in print, and we include the DLA for completeness. In some cases, the errors
cited by \citet{Rao06} may not cover the revised values from higher resolution, higher S/N data, and the plotted range gives an idea of how much things 
may change. However, we find the fits and errors in column density from \citet{Rao06} to be completely satisfactory, and have used them in the body of the
text. In cases where the profiles with error profiles that seem to fit slightly better, there is no systematic sense of the differences: in some cases the
H I column density is smaller, and in others it is larger. This may imply that our values of [Zn/H] may need revision later, when higher quality data become 
available. In particular, the system in Q0826-2230 may have a slightly higher value of N$_{\rm H \ I}$ and therefore slightly lower [Zn/H].
However, the value would still be $\sim$2$\times$ solar. The conclusions of this paper would not be changed by any likely changes in the values of 
N$_{\rm H \ I}$.


\begin{figure}
\includegraphics[height=\linewidth, angle=90]{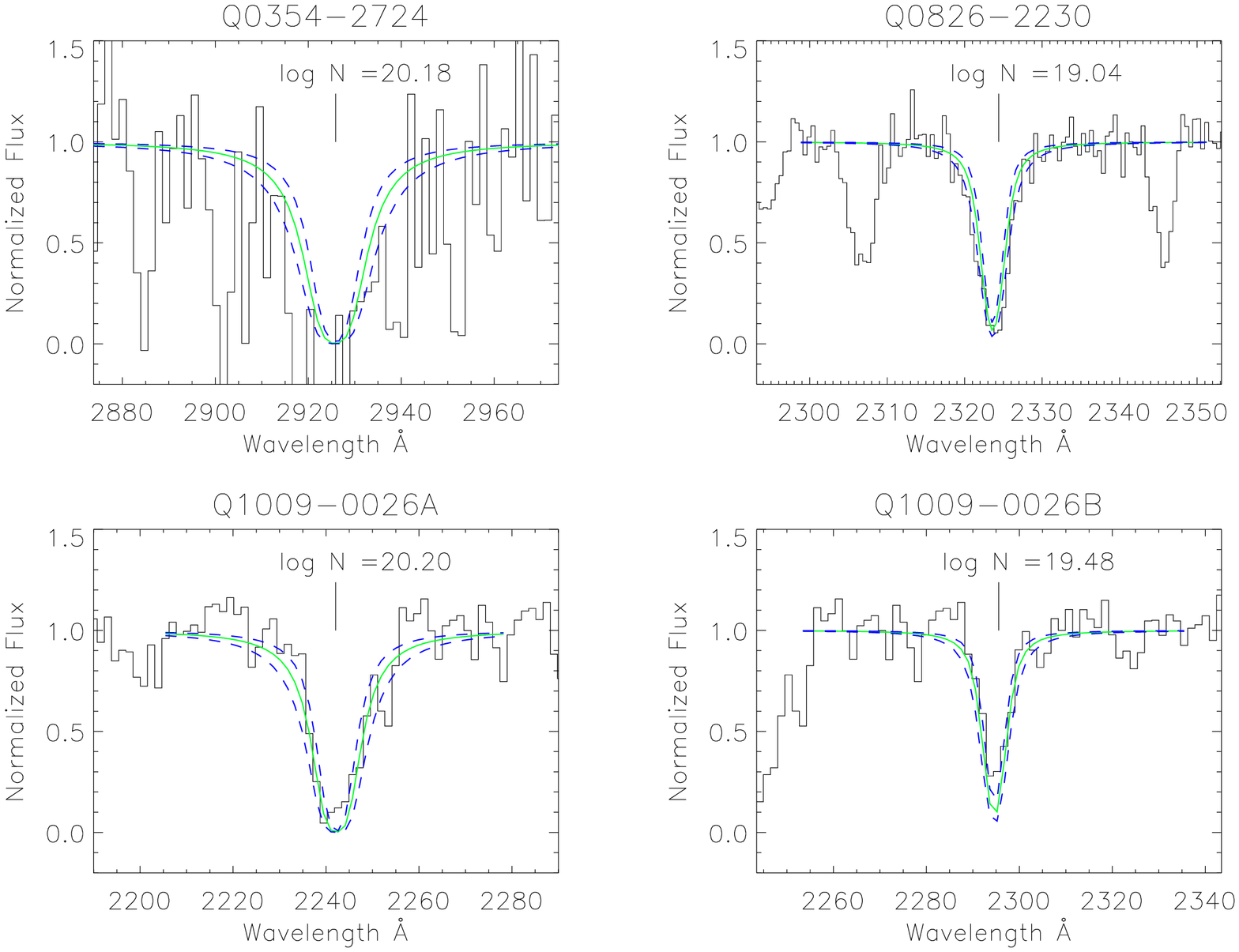}
\caption{UV spectra for the systems, clockwise from the top left; $z_{abs}$=1.4051 system in Q0354-2724, $z_{abs}$=0.9110 in Q0826-2230, 
	$z_{abs}$=0.8866 in Q1009-0026, and $z_{abs}$=0.8426 in Q1009-0026. The central green solid profile in each plot is for the best fit value 
	from \citet{Rao06}, and in the blue dashed profiels above and below, the column density was modified by 0.15 dex. The vertical tick 
	mark above the profile represents the absorption center of the metal lines. Note the different abscissa in the individual figures.}
\end{figure}

\begin{figure}
\includegraphics[height=\linewidth, angle=90]{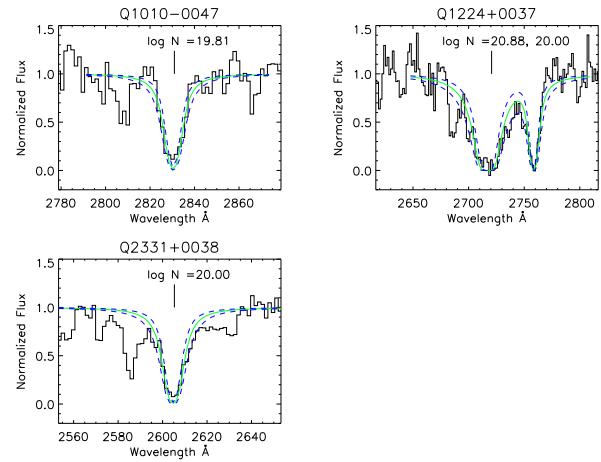}
\caption{Same as figure 14, but for clockwise from the top left; $z_{abs}$=1.3270 in Q1010-0047, $z_{abs}$=1.2346 and $z_{abs}$=1.2665
	in Q1224+0037, and $z_{abs}$=1.1414 in Q2331+0038.}
\end{figure}

\bsp

\label{lastpage}

\end{document}